\documentclass[fleqn,usenatbib]{mnras}

\usepackage{graphicx}
\usepackage{float}
\usepackage{hyperref}
\usepackage{color}
\usepackage{amsmath}
\usepackage{amssymb}
\usepackage{mathrsfs}
\usepackage{epsfig}
\usepackage{epstopdf}
\usepackage{multirow}
\usepackage{longtable}
\usepackage{lscape}
\usepackage[flushleft]{threeparttable}
\usepackage{caption}
\usepackage{gensymb}

\usepackage[T1]{fontenc}
\usepackage{ae,aecompl}

\hypersetup{draft}

\newcommand{\mum}{$\micron$}

\newcommand{\z}{$z$}
\newcommand{\Mstar}{$M_\ast$}
\newcommand{\Mdust}{$M_{\rm dust}$}
\newcommand{\Tdust}{$T$}
\newcommand{\Mgas}{$M_{\rm gas}$}
\newcommand{\Msun}{$M_\odot$}

\newcommand{\Lsun}{$L_\odot$}
\newcommand{\Loiii}{$L_{\rm [OIII]}$}

\newcommand{\TdustCold}{$T_{\rm cold}$}
\newcommand{\TdustWarm}{$T_{\rm warm}$}
\newcommand{\BetaCold}{$\beta_{\rm cold}$}
\newcommand{\BetaWarm}{$\beta_{\rm warm}$}

\newcommand{\Rms}{$R_{\rm MS}$}
\newcommand{\IRAS}{{\it IRAS}}
\newcommand{\WISE}{{\it WISE}}
\title[The cool ISM in radio AGNs]{Quantifying the cool ISM in radio AGNs: evidence for late-time re-triggering by galaxy mergers and interactions}

\author[E.~Bernhard et al.]{E. Bernhard$^{1}$\thanks{E-mail: e.p.bernhard@sheffield.ac.uk},
C. N. Tadhunter$^{1}$,
J. C. S. Pierce$^{1,2}$,
D. Dicken$^{3}$,
J. R. Mullaney$^{1}$,
\newauthor
R. Morganti$^{4}$,
C. Ramos Almeida$^{5,6}$,
and E. Daddi$^{7}$
\\
$^{1}$Department of Physics $\&$ Astronomy, University of Sheffield, Sheffield S3 7RH, UK\\
$^{2}$Centre for Astrophysics Research, University of Hertfordshire, College Lane, Hatfield AL10 9AB, UK\\
$^{3}$Institut d'Astrophysique Spatiale, CNRS, Univ. Paris-Sud, Universit\'{e} Paris Saclay, B\^{a}timent 121, 91405 Orsay Cedex, France\\
$^{4}$ASTRON, the Netherlands Institute for Radio Astronomy, Oude Hoogeveensedijk 4,7991 PD Dwingeloo, The Netherlands\\
$^{5}$Instituto de Astrof\'{i}sica de Canarias, Calle v\'{i}a L\'{a}ctea, s/n, 38205 La Laguna, Tenerife, Spain\\
$^{6}$Departamento de Astrof\'{i}sica, Universidad de La Laguna, 38206, La Laguna, Tenerife, Spain\\
$^{7}$CEA, IRFU, DAp, AIM, Universit\'{e} Paris-Saclay, Universit\'{e} Paris Diderot, Sorbonne Paris Cit\'{e}, CNRS, F-91191 Gif-sur-Yvette, France
}

\begin{document}

\date{Accepted XXX; Received XXX; in original form XXX}

\maketitle

\begin{abstract}
We use deep {\it Herschel} observations of the complete 2Jy sample of powerful radio AGNs in the local universe (0.05~$<$~\z~$<$~0.7) to probe their cool interstellar medium (ISM) contents and star-forming properties, comparing them against other samples of nearby luminous AGNs and quiescent galaxies. This allows us to investigate triggering and feedback mechanisms. We find that the dust masses of the strong-line radio galaxies (SLRGs) in our sample are similar to those of radio-quiet quasars, and that their median dust mass (\Mdust~=~2~$\times~10^{7}~$\Msun) is enhanced by a factor $\sim$200 compared to that of non-AGN ellipticals, but lower by a factor $\sim$16 relative to that of local ultra-luminous infrared galaxies (UILRGs). Along with compelling evidence for merger signatures in optical images, the SLRGs in our sample also show relatively high star-formation efficiencies, despite the fact that many of them fall below the main sequence for star forming galaxies. Together, these results suggest that most of our SLRGs have been re-triggered by late-time mergers that are relatively minor in terms of their gas contents. In comparison with the SLRGs, the radio AGNs with weak optical emission lines (WLRGs) and edge-darkened radio jets (FRIs) have both lower cool ISM masses and star-formation rates (by a factor of $>$30), consistent with being fuelled by a different mechanism (e.g. the direct accretion of hot gas).
\end{abstract}


\section{Introduction}
\label{Intro}

Despite the importance of supermassive black hole growth in regulating galaxy evolution via active galactic nucleus (AGN) feedback (see \citealt{Harrison2017} for a review), AGN triggering mechanisms remain a highly debated subject. In this context, powerful radio AGNs are particularly important. On the one hand, they are almost invariably hosted by elliptical galaxies, allowing for relatively ``clean'' searches to be made for the morphological signatures of the triggering events (e.g. tidal tails and shells); any cool gas detected in such galaxies is also more likely to have an external origin than in late-type galaxies. On the other hand, radio AGNs launch powerful relativistic jets that are capable of heating the large-scale inter-stellar/galactic medium (ISM/IGM) -- one of the most important forms of AGN feedback \citep[e.g.][]{Best2006, McNamara2007}.

Deep observational campaigns undertaken at optical wavelengths have reported that signatures of galaxy interactions are common amongst samples of local powerful radio AGNs \citep[e.g.][]{Heckman1986, Smith1989, RamosAlmeida2011, RamosAlmeida2012, Pierce2021}. However, these optical features do not correspond to a single merger phase (i.e. pre-or-post merger) or type (i.e. minor/major), thereby questioning the link between radio AGN activity and the peaks of major gas-rich galaxy mergers\footnote{Hereafter, we refer to the peaks of major gas-rich galaxy merger to differentiate from galaxy interactions at pre/post galaxy merger and more minor mergers.}, as represented by ultra-luminous infrared galaxies (ULIRGs; \citealt{Sanders1988, Sanders1996}). In addition, a higher fraction of merger signatures were found amongst powerful radio AGNs associated with strong optical emission lines (i.e. strong-line radio galaxies and high-excitation radio galaxies; SLRGs and HERGs), typical of those observed in quasars (QSOs), when compared to radio AGNs lacking such optical features (i.e. weak-line radio galaxies and low-excitation radio galaxies; WLRGs and LERGs; e.g. \citealt{Malin1983, RamosAlmeida2011, Pierce2021})\footnote{In this work, we use the SLRG/WLRG classification, where SLRGs have EW$_{\rm [OIII]}~>~10~\textrm{\AA}$.}. This potentially implies different dominant triggering mechanisms for SLRGs and WLRGs, linking the most major gas-rich mergers with the former population of radio AGNs. However, it is difficult to quantify the nature of the merger (i.e. major/minor and gas-rich/poor) from optical images alone.

With the advancement of mid-to-far-infrared astronomy via observatories such as the Wide-field Infrared Survey Explorer (\WISE; \citealt{Wright2010}), {\it Spitzer} \citep{Werner2004}, and {\it Herschel}\footnote{{\it Herschel} is an ESA space observatory with science instruments provided by European-led Principal Investigator consortia and with important participation from NASA.} \citep{Pilbratt2010}, it is now possible to trace the dust content of galaxies, a proxy for the overall cool interstellar medium (ISM) contents \citep[e.g.][]{Draine2007, Parkin2012, RemyRuyer2014}. This provides information on the nature of the merger by comparing the cool ISM content of powerful radio-loud AGNs with that expected from major gas-rich galaxy mergers, therefore complementing results from optical images. In a preliminary study, \cite{Tadhunter2014} found that the median dust mass of nearby SLRGs is lower than that of ULIRGs, therefore emphasising the importance of more minor mergers in triggering such objects.

The star-forming properties of host galaxies also provide key information on AGN triggering mechanisms. Most star-forming galaxies follow a redshift-dependent relationship between star formation rate (SFR) and stellar mass \cite[e.g.][]{Daddi2007, Elbaz2007, Noeske2007, Rodighiero2014, Sargent2014, Schreiber2015}, known as the main sequence (MS) of galaxies. A small fraction ($\sim$3~per~cent), mainly triggered by major gas-rich galaxy mergers, and referred to as star-bursting systems, show SFRs at least a factor of 4 above the MS \citep[e.g.][]{Sargent2014, Schreiber2015}. Therefore, comparing the star-forming properties of the hosts of powerful radio-loud AGNs to the MS of galaxies can provide further information on their triggering mechanisms.

Here, using deep {\it Herschel} far-infrared (FIR) data for a unique sample of nearby (i.e. \z$~<~$0.7) powerful radio AGNs (the 2Jy sample; see \S\,\ref{2Jy} for details), combined with the abundance of complementary multi-wavelength data available for this sample, we investigate the triggering mechanisms of powerful radio AGNs, split in terms of SLRGs and WLRGs. To do this, we measure their cool ISM properties, SFRs, and SFR efficiencies, and compare them against those of samples of radio-quiet QSOs, non-AGN classical elliptical galaxies, and ULIRGs. We further place the SFRs in the context of the MS of star-forming galaxies to provide a more complete picture of the triggering and feedback mechanisms of powerful radio AGNs.

This paper is organised as follows. We present in \S\,\ref{sec:sampdat} our samples of powerful radio AGNs and comparison samples. We show in \S\,\ref{sec:dustMass}, \S\,\ref{sec:SFRcalc}, and \S\,\ref{sec:Mstarcalc} how we calculated dust masses, SFRs and stellar masses, respectively. Our results on the dust masses and the SFRs are presented in \S\,\ref{sec:results}, and their implications in \S\,\ref{sec:discussion}. Finally, we present the main concluding remarks in \S\,\ref{sec:conclusion}. Throughout, we adopted a WMAP--9 year cosmology ($H_0$~=~69.3~km~s$^{-1}$~Mpc$^{-1}$, $\Omega_m$~=~0.29, $\Omega_{\Lambda}$~=~0.71; \citealt{Hinshaw2013})\footnote{We stress that these are the default values for the WMAP--9 year cosmology of the {\sc python} package ASTROPY \citep{Astropy2018}.} and a \cite{Chabrier2003} initial mass function when calculating galaxy properties.

\section{Data and samples}
\label{sec:sampdat}

In this section we present our samples of powerful radio AGNs and the comparison samples for which we derive dust masses (\Mdust) and SFRs. The main method adopted to calculate \Mdust\ relies on the 100 and 160~\mum\ fluxes and their ratio (see \S\,\ref{sec:dustMass}), since these are now available for large samples of nearby objects. Therefore, our primary selection is based on the availability of {\it Herschel}-PACS \citep{Poglitsch2010} fluxes at 100 or 160~\mum. For objects with additional {\it Herschel}-SPIRE data (i.e. at 250, 350, and 500~\mum; \citealt{Griffin2010}), we constructed infrared (IR) spectral energy distributions (SEDs) to fit detailed dust emission models (see \S\,\ref{subsec:IRSEDfitDustMass}). For these we also included the {\it Herschel} fluxes at 70~\mum\ when available, and we required an additional mid-IR (MIR) data point to account for any potential warmer dust contributions (see \S\,\ref{subsubsec:IRSEDfit}). Overall, sources selected for detailed SED fits fulfilled all of the following criteria as a minimum requirement,

\begin{itemize}
	\item detected with {\it Spitzer} at 24~\mum\ {\it or} \WISE\ at 22~\mum;
	\item detected with {\it Herschel}--PACS at 100~\mum\ or 160~\mum;
	\item detected with {\it Herschel}--SPIRE at 250~\mum\ {\it and} 350~\mum.
\end{itemize}

\noindent These SED fits provide more precise estimates of \Mdust\ values, which, once compared with those estimated from the 100/160~\mum\ flux ratio method (see \S\,\ref{subsec:MdustRatio}), allows us to test the latter.

To calculate SFRs, we used a multi-component SED fitting code which accounts for AGN contribution when necessary (see \S\,\ref{sec:SFRcalc}). Therefore, for these SFR estimates, we also considered archival \WISE, {\it Spitzer}--MIPS \citep{Fazio2004} and {\it Spitzer}--IRAC \citep{Rieke2004} data from the NASA/IPAC IR Science Archive (IRSA) at wavelengths of 8--24~\mum. Only sources with robust quality flags were used\footnote{For the \WISE\ data, where extended sources were flagged, we used the photometry calculated within the extended 2MASS aperture when available. Otherwise, when the extended flag was set to 1, we checked the reduced chi-squared value ($\chi^{2}_{\nu}$) of the profile fit, and only used the flux if $\chi^{2}_{\nu}~<~3$. For the $Spitzer$--MIPS and --IRAC data, only point sources were used, and the fluxes of extended sources were discarded.}, as described in the explanatory supplements of each instrument available on the IRSA website\footnote{Available at \url{https://wise2.ipac.caltech.edu/docs/release/allwise/expsup/index.html} for \WISE, and at \url{https://irsa.ipac.caltech.edu/data/SPITZER/Enhanced/SEIP/docs/seip_explanatory_supplement_v3.pdf} for $Spitzer$.}.

We provide in Table\,\ref{tab:sample} a full listing of our various populations of AGNs and galaxies, as presented in the following subsections. In addition, detailed Tables listing the general properties of each of the objects in these samples are accessible in the online material.

\setlength{\tabcolsep}{3pt}
\begin{table}
\begin{threeparttable}
\caption{Table summarising our various populations and samples of AGNs and galaxies, as described in \S\,\ref{sec:sampdat}. Here, ``RL'', ``RQ'' and ``E.'' correspond to radio-loud, radio-quiet, and ellipticals. We also indicate where the {\it Herschel} IR fluxes were taken from (i.e. ``C14'': \citealt{Ciesla2014}, ``W16'': \citealt{Westhues2016}, ``S18'': \citealt{Shangguan2018}, ``S19'': \citealt{Shangguan2019}, ``D21'': Dicken et al., in prep., and/or HPDPs), as well as some sample statistics: the total number of sources retained in each sample ($N_{\rm tot}$), of which the number selected for SED fits ($N_{\rm fit}$; see \S\,\ref{sec:sampdat}), and the fractions detected at both 100 and 160~\mum, as well as the fractions detected at one band only, respectively. We note that the large fractions of detected sources in the Atlas$^{3 \rm D}$ sample arise due to selection effects, and do not reflect the true fractions of IR detected elliptical galaxies in this sample. The K17$^{*}$ corresponds to the extra sources with dust masses reported in \protect\cite{Kokusho2019} and based on AKARI fluxes measured in \protect\cite{Kokusho2017} for the Atlas$^{3 \rm D}$ sample, and that we have included in our analysis (see \S\,\ref{atlas3d}). \label{tab:sample}}
\centering
\begin{tabular}{ccccccc}
\hline
\multirow{2}{*}{Types} &  \multirow{2}{*}{Samp.}  & \multirow{2}{*}{Ref. IR} & \multirow{2}{*}{$N_{\rm tot}$} & \multirow{2}{*}{$N_{\rm fit}$} & \multicolumn{2}{c}{IR det.} \\
 &   &  &  &  & 2 bands & 1 band  \\
\hline
\multirow{2}{*}{RL AGNs}     & 2Jy                                 & D21      &  46                 & 8                   & 89\%                   & 11\%  \\\vspace{10pt}
                             & 3CR                                 & W16      &  45                 & 4                   & 47\%                   & 29\% \\
\multirow{2}{*}{RQ QSOs}     & PGQs                                & S18      &  70                 & 9                   & 86\%                   & 10\% \\\vspace{10pt}
                             & Type-II                             & S19      &  86                 & 12                  & 82\%                   & 13\% \\
\multirow{2}{*}{ULIRGs}      & \multirow{2}{*}{HERUS}              & C18      & \multirow{2}{*}{41} & \multirow{2}{*}{14} & \multirow{2}{*}{51\%}  &  \multirow{2}{*}{49\%} \\\vspace{10pt}
                             &                                     & HPDPs    &                     &                     &                        &  \\
\multirow{4}{*}{E. Gal.}     & \multirow{2}{*}{Atlas$^{3 \rm D}$}  & HPDPs    & 6                   & 4                   & 67\%                   &  33\%    \\\vspace{5pt}
                             &                                     & K17$^*$  & 32                  & --                  & --                     &  \\
                             & \multirow{2}{*}{HRS}                & C14      & \multirow{2}{*}{8}  & \multirow{2}{*}{0 } & \multirow{2}{*}{0\%}  & \multirow{2}{*}{0\%}     \\
                             &                                     & HPDPs    &                     &                     &                        &  \\
\hline
\end{tabular}%
\end{threeparttable}
\end{table}

\subsection{Powerful radio-loud AGNs}
\label{subsec:radioGal}

Our main samples consist of nearby radio-loud AGNs. As they are almost invariably hosted in elliptical galaxies \citep[e.g.][]{Tadhunter2016, Pierce2021}, they offer a clean way to search for signs of interactions and investigate triggering and feedback mechanisms (see \S\,\ref{Intro}). 

\label{subsec:samp_radgal}
\subsubsection{The 2Jy sample}
\label{2Jy}

Our primary radio AGN sample comprises 46 nearby (0.05$~<~$\z$~<~$0.7) southern ($\delta~<~10{\degree}$) powerful radio galaxies with steep radio spectra ($F_\nu~\propto~\nu^{-\alpha}$; $\alpha~>~0.5$) from the 2Jy sample of \cite{Wall1985}, complete at $S_{\rm 2.7GHz}~>~2$~Jy \citep{Dicken2009}. This sample is unique in the depth and completeness of its multi-wavelength data, in particular at mid-to-far-IR wavelengths. Deep {\it Herschel} photometry is now available for the 2Jy sample, following observations performed between October 2012 and March 2013 as part of program DDT\_mustdo\_4. While preliminary results for the {\it Herschel} data were presented in \cite{Tadhunter2014}, the detailed observation strategy, data reduction, and the most recent fluxes will be reported in Dicken et al. (subm.). Crucially, in the latter, efforts were made to calculate IR fluxes free of non-thermal contamination, which can be a particular issue for radio-loud AGNs in the FIR.

All of the 46 sources were detected at 100~\mum, and 41 (89~per~cent) were also detected at 160~\mum. However, we treated the FIR fluxes of 14 sources (30~per~cent) as upper limits, since they showed potential non-thermal contamination that could not be corrected for (Dicken et al., subm.; see also Tables in the online material). Out of the full sample of 46 sources, eight were selected for detailed SED fits based on the criteria outlined at the beginning of \S\,\ref{sec:sampdat}.

For this sample, we also have radio morphologies separated between FRIs and FRIIs \citep[following][]{Fanaroff1974} based on the radio observations of \cite{Maroganti1993, Maroganti1999}, [OIII]$_{\lambda_{5007\textrm{\AA}}}$ luminosities (\Loiii) and optical classes (i.e. SLRG/WLRGs) based on the optical spectroscopic data of \cite{Tadhunter1993, Tadhunter1998}. Preliminary results on the dust masses of the SLRGs in this sample were presented in \cite{Tadhunter2014}.

\subsubsection{The 3CR sample}
\label{3CR}

We further considered the Revised Third Cambridge Catalogue of Radio Sources \citep[][3CR]{Bennett1962a, Bennett1962b, Spinrad1985}, which is a flux-limited sample of bright ($S_{\rm 178Mhz}~>~9$~Jy) northern ($\delta~>~-5{\degree}$) radio AGNs. The 3CR sample is similar to the 2Jy sample in terms of radio power, but selected at lower frequencies. Although less complete in terms of detections at FIR wavelengths (see below), it provides an important check on the results for the 2Jy sample, using a sample of powerful radio AGNs selected at a different frequency. A representative sub-sample of 48 3CR sources at \z$~<~$0.5 have been observed with {\it Herschel}, and the fluxes and upper limits are reported in Table~4 of \cite{Westhues2016}.

We removed 3C~459.0 as it is the same object as PKS~2314+03 from the 2Jy sample (see \S\,\ref{2Jy}). Out of the 47 unique remaining 3CR sources, 29 (62~per~cent) were detected at 100~\mum, of which 22 (47~per~cent) were also detected at 160~\mum. The {\it Herschel} observations of the 3CR sample are shallower than those of the 2Jy, hence the lower detection rates. Including upper limits, we found 41 sources (87~per cent) with fluxes or upper limits at 100 or 160~\mum. For an extra four sources without {\it Herschel}--PACS fluxes in \cite{Westhues2016}, we found archival Infrared Astronomical Satellite (\IRAS) fluxes at 100~\mum\ from NED, one of which is an upper limit. Therefore, out of our sample of 47 unique 3CR sources, 45 were retained.

The possibility of non-thermal contamination was addressed by collecting archival measurements at $\lambda~>~500~$\mum\ (e.g. from the Very Large Array, the Very Long Baseline Array, and the IRAM 30-meter telescope) from the NASA/IPAC Extragalactic Database (NED), separating the core and the extended emission (lobe/hotspot), when possible. A power-law with free spectral index was fit to the extended component, and extrapolated down to 100~\mum. For the core, we took the level of non-thermal flux found at longer wavelengths and extrapolated this down to 100~\mum, assuming a flat spectrum. If the extrapolated fluxes (i.e. from the extended and/or core emission) at any of the {\it Herschel} wavelengths were above 10~per~cent of the observed fluxes, the source was considered as potentially contaminated by non-thermal emission. However, for sources showing apparent non-thermal contamination from the extended radio lobes only, we compared the radio maps against the size of the {\it Herschel} beams. If the {\it Herschel} beams were smaller than the bulk of extended non-thermal emission, the fluxes were considered free of non-thermal contamination.

The FIR fluxes contaminated by non-thermal emission, either from the core and/or for which the {\it Herschel} beams potentially contained the extended non-thermal emission, were treated as upper limits. Amongst the 45 radio AGNs in the 3CR sample, the FIR fluxes for six objects (13~per~cent of the full 3CR sample and 24~per~cent of the detected sources) were treated as upper limits to account for the potential non-thermal contamination. Overall, four 3CR sources were selected for detailed SED fits based on the criteria outlined at the beginning of \S\,\ref{sec:sampdat}, one of which includes the \IRAS\ flux at 100~\mum.

The 3CR objects at \z$~\lesssim~$0.3 were originally classified as LERGs or HERGs by \cite{Buttiglione2009, Buttiglione2010, Buttiglione2011}, but then re-classified as SLRGs or WLRGs by \cite{Tadhunter2016}. Although LERG/HERG and WLRG/SLRG classification schemes show considerable overlap, they are not exactly the same \citep[see discussion in][]{Tadhunter2016}. For this work, we use the WLRG/SLRG classifications of the 3CR sources, for which we also have radio classes (FRI/FRII) and [OIII] luminosities. We also found archival\footnote{\cite{Leahy1986}, \cite{Leahy1991}, \cite{Gelderman1994}, \cite{Giovannini1994}, \cite{Jackson1997}, \cite{Mack1997}, \cite{Ludke1998}, \cite{Haas2005}, \cite{Koss2017}, R.A. Laing (unpublished), and D.A. Clarke and J.O. Burns (unpublished).} radio classes and \Loiii\ values for all of the remaining 3CR sources in \cite{Westhues2016} with \z$~\gtrsim~$0.3.

\subsection{Radio-quiet QSOs}
\label{subsec:QSO}

To test whether the cool ISM and star-forming properties of radio-loud and radio-quiet AGNs are different, perhaps related to distinct triggering mechanisms, we also constructed comparison samples of optically unobscured (Type-I) and obscured (Type-II) radio-quiet QSOs.

\label{subsec:samp_QSO}
\subsubsection{The Type-I PG QSO sample}
\label{subsec:samp_pgqs}

For our comparison sample of Type-I QSOs, we used the 87 nearby (\z$~<~$0.5) UV/optically selected QSOs from the Palomar-Green (PG) survey of \cite{Schmidt1983} with $B~<~16.17$ (e.g. \citealt{Goldschmidt1992}). The PG QSO (PGQ) sample is representative of bright, nearby Type-I (unobscured) QSOs and benefits from a plethora of multi-wavelength data.

After removing the 16 radio-loud PGQs \citep[as classified according to the criteria of][]{Boroson1992}, we were left with 71 objects. We adopted the {\it Herschel} fluxes listed in \citeauthor{Shangguan2018} (\citeyear{Shangguan2018}; and references therein), and found 68 sources (96~per~cent) detected at 100~\mum, of which 60 (84~per~cent) were also detected at 160~\mum. For all of the remaining sources, but one not observed with {\it Herschel}--PACS, we used the upper limits as reported in \cite{Shangguan2018}. Therefore, our final PGQ sample contained 70 sources, of which nine were selected for detailed SED fits based on the criteria outlined at the beginning of \S\,\ref{sec:sampdat}. For the full PGQ sample we have [OIII] luminosities, and estimates of the dust and stellar masses \citep{Shangguan2018}.

\subsubsection{The Type-II QSO sample}
\label{subsec:samp_t2qsos}

For our comparison sample of Type-II QSOs, we used that defined in \cite{Shangguan2019}, and originally taken from the Sloan Digital Sky Survey \citep[SDSS;][]{Reyes2008}. It contains 86 randomly selected sources that match the PGQ sample of \cite{Shangguan2018}, in redshift (at \z$~<~$0.5) and [OIII] luminosity \citep[$10^{8.0}~<~$\Loiii/\Lsun$~<~10^{9.8}$; see][for details on the selection technique]{Shangguan2019}. We used the {\it Herschel} fluxes reported in \cite{Shangguan2019}, and found 82 sources (95~per~cent) detected at 100~\mum, of which 71 (82~per~cent) were also detected at 160~\mum. For the remaining sources we used the upper limits on the {\it Herschel}--PACS fluxes reported in \cite{Shangguan2019}. This sample is representative of bright, nearby Type-II QSOs.

According to \cite{Shangguan2019}, none of these sources were found to show significant non-thermal contamination at IR wavelengths. As for the PGQs, we also have [OIII] luminosities, and dust and stellar mass estimates for the Type-II QSOs \cite[see][for details]{Shangguan2019}. Out of the 86 objects, 12 were selected for detailed SED fits based on the criteria outlined at the beginning of \S\,\ref{sec:sampdat}.

\subsection{Ultra Luminous IR Galaxies}
\label{subsec:samp_ulirgs}

We defined a comparison sample of ULIRGs which traces the cool ISM content of galaxies at the peaks of major gas-rich mergers. This allows us to assess the importance of major gas-rich mergers for triggering the powerful radio AGNs in our samples, by comparing their cool ISM properties against those of ULIRGs\footnote{We stress that, although the majority of ULIRGs are likely to represent the peaks of major, gas-rich mergers, not all such mergers necessarily lead to the levels of star formation and AGN activity observed in ULIRGs. However, all major, gas-rich mergers would be expected to have substantial (ULIRG-like) reservoirs of cool ISM.}.

We used the {\it Herschel} ULIRG survey (HERUS, PI D. Farrah, programme ID OT1\_dfarrah\_1), which is an unbiased sample of 43 nearby (\z$~<~0.3$) ULIRGs with \IRAS\ 60~\mum\ fluxes $>$~1.8~Jy, and originally identified in the \IRAS\ Point Source Catalogue Redshift (PSC-z) survey of \cite{Saunders2000}. The ULIRGs Mrk~1014 and 3C~273.0 were removed from this sample, as they correspond to AGNs found in our PGQ (\S\,\ref{subsec:samp_pgqs}) and 3CR (\S\,\ref{3CR}) samples, respectively.

For the {\it Herschel}--PACS data, we cross-matched the ULIRG sample with archival {\it Herschel} data from the Highly Processed Data Products (HPDPs), available on the IRSA website, which offer the most complete and uniform database of reduced fluxes for {\it Herschel}--PACS \citep{Marton2017} and {\it Herschel}--SPIRE \citep{Schulz2017}\footnote{As found in \cite{Bernhard2021}, some nearby sources can be misclassified as point sources in the HPDPs, systematically under-estimating their fluxes. For these, we have corrected the fluxes for missed extended emission, as described in Appendix A of \cite{Bernhard2021}.}. We found 19 sources (46~per~cent) observed and detected at 100 and 160~\mum. For all of the remaining sources (54~per~cent), we used the original \IRAS\ fluxes at 100~\mum\ (and 60~\mum\ for the SED fits). This sample was also observed by {\it Herschel}--SPIRE, with all of the 41 sources being detected, and with fluxes reported in the Table~2 of \citeauthor{Clements2018} (\citeyear{Clements2018}; originally from \citealt{Pearson2016}). Finally, 14 of the ULIRGs were selected for detailed SED fits based on the criteria outlined at the beginning of \S\,\ref{sec:sampdat}.

\subsection{Classical elliptical galaxies}
\label{subsec:samp_ellip}

As powerful radio AGNs are almost invariably hosted by elliptical galaxies, we defined samples of non-AGN classical elliptical galaxies to compare against.

\subsubsection{The Atlas$^{\rm 3D}$ sample}
\label{atlas3d}

Our first sample was selected from the Atlas$^{3\rm D}$ survey, which is a volume limited (D$~<~$42~Mpc, $M_{K}~<~-21.5$) sample of 260 nearby morphologically-selected early-type galaxies \citep{Cappelliari2011}. We only retained the 68 objects that were classed as elliptical (i.e. T-Type~$\lesssim$~-3.5 in \citealt{Cappelliari2011}, excluding S0 galaxies) to match the morphology of the hosts of powerful radio AGNs.

We cross-matched this sample against archival {\it Herschel} data from the HPDPs (see \S\,\ref{subsec:samp_ulirgs} for the HPDPs), and, out of the 25 objects observed with {\it Herschel}, six (30~per~cent) had reliable detected fluxes at 100~\mum, of which four (20~per~cent) were also detected at 160~\mum. In total, seven galaxies were at least detected at 100 or 160~\mum. We removed NGC~4374 as it showed an inflection at {\it Herschel}-SPIRE wavelengths, which is a potential sign of non-thermal AGN contamination.

Therefore, our full sample of elliptical galaxies from the Atlas$^{3\rm D}$ survey contained six objects, of which four were selected for detailed SED fits based on the criteria outlined at the beginning of \S\,\ref{sec:sampdat}. We note that, as a consequence of using only elliptical galaxies detected at FIR wavelengths, this sample is likely biased toward the nearby non-AGN elliptical galaxies that are brightest at FIR wavelengths.

The Atlas$^{3\rm D}$ survey further benefits from IR observations taken with AKARI at 9, 18, 65, 90, and 140~\mum\ \citep{Kokusho2017}. Using the latter, dust masses were calculated in \cite{Kokusho2019}. We benefited from the agreement between the dust masses measured with {\it Herschel} and AKARI (see \S\,\ref{subsubsec:fitCompPast}) to significantly increase the size of our sample of non-AGN elliptical galaxies. To do this, we included an extra 32 elliptical galaxies (i.e. excluding S0) that were not observed by {\it Herschel}, but had measured dust masses (11 as upper limits) derived from the AKARI observations and listed in \cite{Kokusho2019}.\footnote{We note that, originally, there were 39 elliptical galaxies with measured \Mdust\ in \cite{Kokusho2019}, including the 11 upper limits, that could have been included in our sample since not observed with {\it Herschel}. However, seven sources with measured dust masses in \cite{Kokusho2019} appeared with fluxes that were not detected (i.e. $<1\sigma$) at any of the AKARI wavelengths in \cite{Kokusho2017}. This led to unphysical values for the dust masses of these seven objects. Because the reasons behind this discrepancy are unclear, we have excluded these objects from our sample. We only retained objects that were at least detected in one AKARI band (at $>2\sigma$).} We note that the AKARI data are much less sensitive than the {\it Herschel} observations. However, they are more complete in their coverage of the Atlas$^{3\rm D}$ sample.

\subsubsection{The {\it Herschel} Reference Survey sample}
\label{hrs}

To further increase the statistics of our sample of classical elliptical galaxies, we used the Herschel Reference Survey (HRS). This is a volume limited sample (i.e. 15$~\lesssim~D~\lesssim~$25~Mpc) of 323 galaxies \citep{Boselli2010}, of which 62 are early-types with 2MASS $K$-band magnitudes $K_{\rm S}~\leq~8.7$~mag, after the revised classification of \cite{Smith2012b}. We only retained the nine objects which are classified as classical ellipticals, and which do not show any signs of AGN activity, as reported in \cite{Smith2012b}, and references therein.

Using the fluxes taken from \cite{Smith2012b}, out of these nine sources, three (33~per~cent) were detected with \IRAS\ at 100~\mum, and the rest had flux upper limits, either from \IRAS\ (1 source) or from {\it Herschel} (5 sources). Two sources were also detected at 160~\mum\ with {\it Herschel}, and five had upper limits. We cross-matched this sample with the {\it Herschel}-SPIRE fluxes presented in \cite{Ciesla2014}, and found that all were constrained by flux upper limits only at longer FIR wavelengths. Therefore, none were selected for detailed SED fits based on the criteria outlined at the beginning of \S\,\ref{sec:sampdat}. In addition, we have removed NGC~4649 from the sample as we found large systematic offsets between the \IRAS\ flux at 100~\mum\ and the {\it Herschel} fluxes, suggesting a flux calibration error. We were therefore left with eight sources, mostly constrained by upper limits alone.

\section{Measuring dust masses}
\label{sec:dustMass}

In this work, we aim to probe the cool ISM content of powerful radio AGNs by measuring their dust masses, and comparing against those of radio-quiet QSOs, ULIRGs, and non-AGN classical elliptical galaxies (see \S\,\ref{sec:sampdat} for the samples).

The dust masses of the galaxies were calculated using \citep[e.g.][]{Mattsson2015},

\begin{equation}
\label{eq:dustMass}
M_{\rm dust} = \frac{{S_{160}~d^2}}{\kappa_{160}~B_{\nu}(160,T)~(1+z)}~M_{\odot},
\end{equation}

\noindent where $S_{160}$ is the measured flux at the wavelength that 160~\mum\ is shifted to in the observer's frame, $d$ is the luminosity distance, $\kappa_{160}$ is the opacity of the dust grains at 160~\mum, and $B_{\nu}(160,T)$ is the specific intensity of a black-body curve evaluated at 160~\mum, and for a given dust temperature, \Tdust.\footnote{We stress that we use the 160~\mum\ flux, since 160~\mum\ is the longest wavelength with a measured flux available for most of our objects, most of which lack {\sc SPIRE} measurements.} $B_{\nu}(160,T)$ requires the determination of the temperature of the dust. To do this, we employed two different methods. For the sub-sample of objects that were selected for detailed SED fits (based on the criteria outlined in \S\,\ref{sec:sampdat}), $B_{\nu}(160,T)$ was measured after performing full IR SED fits, as fully described in \S\,\ref{subsec:IRSEDfitDustMass}. For the rest of our objects, we developed a method to calculate \Tdust\ mostly based on the 100/160~\mum\ flux ratios, and following that used in \cite{Tadhunter2014}, but calibrated using our \Mdust\ measured from the detailed SED fits, as fully described in \S\,\ref{subsec:MdustRatio}. We stress that, for consistency, we will use values of \Mdust\ measured from the 100/160~\mum\ flux ratios across all of our samples while comparing their cool ISM content, and IR SED fits were only used to test the flux ratio technique.

We also assumed a value for the opacity of the dust grains at 160~\mum, $\kappa_{160}$ (Eq.\,\ref{eq:dustMass}), a parameter which is known to be highly uncertain (see Fig.\,1 of \citealt{Clark2016} for different values of the dust grain opacities across various studies). The values of $\kappa_{160}$ that are often quoted in studies of galaxies \citep[e.g.][]{James2002, Draine2003, Clark2016} differ by up to a factor of 3.5. For consistency, we adopted a single value of the dust opacity throughout ($\kappa_{160}=1.038~{\rm m^2~kg^{-1}}$ from \citealt{Draine2003}), which is identical to that used in our various comparison studies (see \S\,\ref{subsubsec:fitCompPast}).

\subsection{The dust masses from SED fits}
\label{subsec:IRSEDfitDustMass}

\subsubsection{Infrared SED fits}
\label{subsubsec:IRSEDfit}

To first order, the thermal dust emission of galaxies can be represented by a modified black-body curve with temperature \Tdust\ and beta index $\beta$ \citep[e.g.][]{Galliano2018}:

\begin{equation}
S_{\nu} (\nu, T, \beta) = {A_{\rm norm}}\times\nu^{\beta}\times{B_{\nu}(\nu,T)},
\label{eq:modBB}
\end{equation}

\noindent where ${S_{\nu}}$ is the flux density, ${B_{\nu}(\nu, T)}$ is the specific intensity of a black-body curve at temperature \Tdust, and $A_{\rm norm}$ is a constant of normalisation. However, the full IR emission of galaxies is a mixture of dust at different temperatures, and attempting to model these with a black-body curve at a single temperature biases the inferred properties of the cooler dust component \citep[e.g.][]{Juvela2012, Hunt2015}. This effect is enhanced in the presence of an AGN, since it is able to heat dust at temperatures typically corresponding to the near-to-mid-IR regime, and with some evidence of FIR emission \citep[e.g.][Dicken et al. subm.]{Dicken2009, Mullaney2011, Siebenmorgen2015, Symeonidis2017, Bernhard2021}.

\begin{figure}
\begin{center}
\includegraphics[width = 0.47\textwidth]{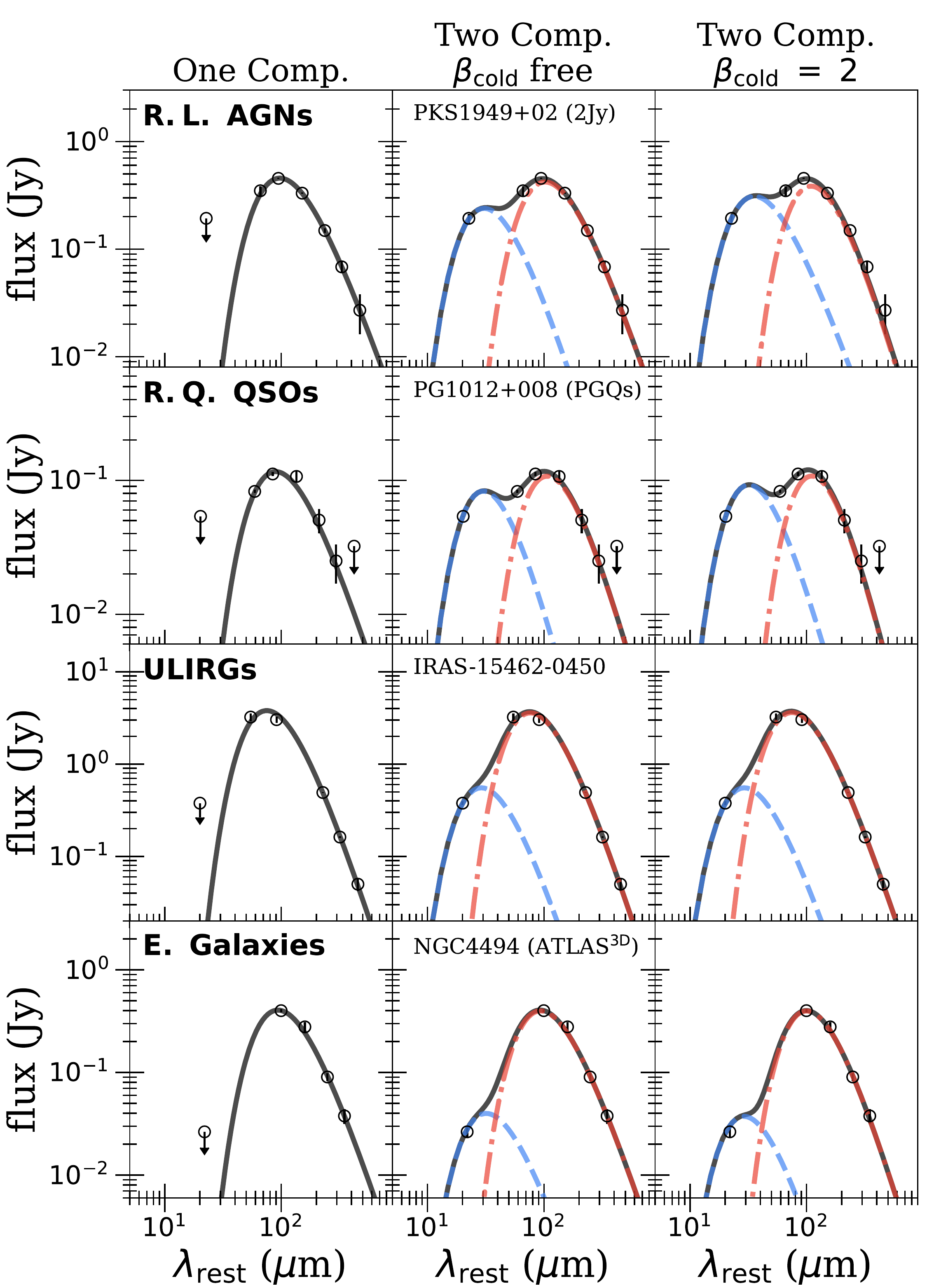}
\caption{Examples of SED fits for one objects of each of our populations of galaxies. From top-to-bottom, these are radio-loud AGNs, radio-quiet QSOs, ULIRGs, and non-AGN elliptical galaxies, as indicated in the top-left corner of each of the leftmost panels. The names of the sources and their respective samples selected to illustrate our SED fits are indicated at the top of each of the central panels. From left-to-right, the panels correspond to each of our three different models for the dust emission as described in \S\,\ref{subsubsec:IRSEDfit}. The observed fluxes are shown with open circles, and the best SED fits with a continuous black line. Downward arrows indicate upper limits on the fluxes. The warmer and the colder dust contributions are shown with a dashed blue line, and a dot-dashed red line, respectively, for corresponding models.$^{\ref{refTableFits}}$}\label{fig:exFit}
\end{center}
\end{figure}

To fit our IR SEDs we tested several models. Our first model is a single modified black-body curve, for which \Tdust\ and $\beta$ were free to change (see leftmost panels in Fig\,\ref{fig:exFit}). In fact, these models are still used in the literature to fit galaxy IR SEDs, probably due to their simplicity \cite[e.g.][]{Clements2018}. Our second model was a combination of two modified black-body curves, defined with two dust temperatures and beta indices (\TdustCold, \BetaCold\ and \TdustWarm, \BetaWarm\ for the cold and the warm dust, respectively; see central panels in Fig\,\ref{fig:exFit}). All of these parameters were free to change. Our third model consisted of two modified black-body curves, with temperatures \TdustCold\ and \TdustWarm, but with \BetaCold~=~2 fixed, which is a common value adopted in studies of galaxies \citep[e.g.][see rightmost panels in Fig\,\ref{fig:exFit}]{Dunne2001, Vlahakis2005, Smith2012b, Cortese2014}. Therefore, we had in total three different models for the IR emission of our galaxies, two of which used a combination of two modified black-body curves at different temperatures (see Fig.\,\ref{fig:exFit}).

We performed maximum likelihood estimation (MLE) to optimise the free parameters of each of these models, and fit the SEDs. We used MLE as it allowed us to easily consider upper limits and errors on the fluxes in a self-consistent way \citep[e.g.][]{Bernhard2019, Grimmett2019}. Due to the complexity of our likelihood function, it could not be maximised analytically. Instead, we maximised it by randomly sampling the posterior distributions of our free parameters, employing the affine invariant ensemble sampler of \cite{Goodman2010}, fully implemented into {\sc emcee}\footnote{{\sc emcee} is publicly available at \url{http://dfm.io/emcee/current/}} \citep{Foreman2013}. The benefit was that we obtained best fitting values with meaningful uncertainties that fully accounted for the presence of upper limits. The median values of the posterior distributions were taken as best fit parameters. Their 1$\sigma$ uncertainties were estimated by using the standard deviations of the posterior distributions, taking into account the covariance between different parameters (e.g. the \TdustCold-\BetaCold\ anti-correlation, in particular).

To reach convergence faster and avoid degeneracies, we reduced the parameter space to physically meaningful values. To do this, we used bounded, normally-distributed priors for each of the parameters defining our models. The priors were such that the explored parameter space was largely consistent with parameters reported in studies of star-forming galaxies \citep[e.g.][]{Hunt2015, Orellana2017}, as well as those including AGN contributions \citep[e.g.][]{Tadhunter2014}. While attempting to fit the IR SEDs of our samples of ULIRGs and non-AGN elliptical galaxies with our two-component models, we found some degeneracies when considering the warmer and colder $\beta$ indices independently. Therefore, we assumed \BetaCold~=~\BetaWarm\ when fitting these samples, since both Rayleigh-Jeans tails of the warmer and cooler dust contributions are expected to arise from star formation. In contrast, \BetaCold\ and \BetaWarm\ were kept independent while fitting the IR SEDs of radio-loud AGNs and radio-quiet QSOs to account for potential differences of the dust properties between those of extended star-forming regions and those of the compact nuclear regions \citep[e.g.][]{Siebenmorgen2015}. 

We show in Fig.\,\ref{fig:exFit} example SED fits for one object in each of our populations of galaxies, and fit with each of our three models for the dust emission.\footnote{The full sets of SEDs and best fitting parameters are available in the online material.\label{refTableFits}} When using our model with a single black-body curve (see leftmost panels in Fig.\,\ref{fig:exFit}), we treated the fluxes at $\lambda~<~$60~\mum\ as upper limits since the model was not designed to represent the full IR emission of galaxies, where the contribution of the warmer dust can be significant at shorter IR wavelengths. The mean \BetaCold\ and \TdustCold\ values and typical range of each of our samples and models are listed in Table~\,\ref{tab:resFit}.

\subsubsection{Results of the detailed fits}
\label{subsubsec:fitresults_new}

We first note that each of our three models provide a good fit to the IR SEDs of our samples, whether hosting an AGN or not (see Fig.\,\ref{fig:exFit} and online material). Consistent with previous work (see \citealt{Galliano2018} for a review), we find that employing a single modified black-body curve generally leads to higher temperatures, and lower $\beta$ indices for the cooler dust, when compared to employing two modified black-body curves (see Table~\,\ref{tab:resFit}). We note, however, that this is not true for our sample of non-AGN elliptical galaxies, where no differences are found for the mean $\beta$ indices and temperatures of the two models. In fact, elliptical galaxies are likely to contain less warmer dust (i.e. evolved galaxies with lower star formation), when compared to our other samples. Therefore, they are equally well represented by a single black-body curve. We also find that the differences on the mean parameters for the cooler dust, between the one and two-component black body models for the ULIRGs, are less significant, when compared to those found for our samples of AGNs (see Table~\,\ref{tab:resFit}). This is likely related to the presence of a hotter, more prevalent, AGN-heated dust contribution in the latter.

For our models with two modified black bodies, fixing \BetaCold~=~2 leads to mean \TdustCold\ values that are systematically reduced by $\sim$2-to-6~K, when compared to models with \BetaCold\ unconstrained. We further note that, when unconstrained, our \BetaCold\ indices are systematically lower than the value of 2, which is the value often used in studies of galaxies (see Table~\,\ref{tab:resFit}). Overall, the ranges of mean \TdustCold\ (i.e. $\sim$25--40~K) and \BetaCold\ (i.e. $\sim$1.4--1.9) values found for our model with two black body components are consistent with those reported in studies of star-forming galaxies \citep[e.g.][]{Hunt2015, Orellana2017}.

\setlength{\tabcolsep}{4pt}
\begin{table*}
\begin{threeparttable}
\caption{List of the mean averages and typical ranges for the best fitting parameters for our galaxies with SEDs that could be fit, combined in terms of galaxy populations and samples (rows in Table), and for each of our three models for the emission of the dust, as described in \S\,\ref{subsubsec:IRSEDfit} (columns in Table). The names of the models are ``One comp.'', ``Two comp.'', and ``Two comp. $\beta_{\rm cold}$~=~2'', and correspond to our single black-body curve, two-component black-body curve, and two-component black-body curve with $\beta_{\rm cold}$~=~2 fixed, respectively. The census for each of the samples is indicated between brackets under the sample names (see also Table~\,\ref{tab:sample}). \label{tab:resFit}}

\centering
\begin{tabular}{cccccccccc}
\hline
Populations &  Samples  & \multicolumn{2}{c}{One comp.} & & \multicolumn{2}{c}{Two comp.} & &\multicolumn{2}{c}{Two comp. $\beta_{\rm cold}~=~2$} \\
     &           &  $\beta_{\rm cold}$ &  $T_{\rm cold}~$(K) & & $\beta_{\rm cold}$ &  $T_{\rm cold}~$(K) & & $\beta_{\rm cold}$ &  $T_{\rm cold}~$(K) \\
\hline
\multirow{4}{*}{Radio AGNs} & 2Jy & \multirow{2}{*}{1.2 (0.3)} & \multirow{2}{*}{36.9 (4.2)} & & \multirow{2}{*}{1.5 (0.2)} & \multirow{2}{*}{32.8 (3.8)} & & \multirow{2}{*}{2.0} & \multirow{2}{*}{28.0 (3.7)} \\\vspace{5pt}
                      &  (8 sources)   & &  & &  &  & &  \\
                      & 3CR & \multirow{2}{*}{1.4 (0.6)} & \multirow{2}{*}{29.2 (9.5)} & & \multirow{2}{*}{1.7 (0.2)} & \multirow{2}{*}{24.7 (4.8)} & & \multirow{2}{*}{2.0} & \multirow{2}{*}{22.7 (4.2)} \\\vspace{10pt}
                      &  (4 sources)  & &  & & &  & &  \\
\multirow{4}{*}{QSOs} & PGQs & \multirow{2}{*}{0.7 (0.2)} & \multirow{2}{*}{40.9 (4.9)} & & \multirow{2}{*}{1.5 (0.2)} & \multirow{2}{*}{27.1 (3.8)} & & \multirow{2}{*}{2.0} & \multirow{2}{*}{23.6 (2.9)} \\\vspace{5pt}
                      &  (9 sources)  & &  & & &  & &  \\
                      & Type-II & \multirow{2}{*}{0.8 (0.3)} & \multirow{2}{*}{44.6 (6.2)} & & \multirow{2}{*}{1.4 (0.2)} & \multirow{2}{*}{30.1 (9.1)} &  &\multirow{2}{*}{2.0} & \multirow{2}{*}{23.8 (4.8)} \\\vspace{10pt}
                      &  (12 sources)  & &  & & &  & &  \\
\multirow{2}{*}{ULIRGs} & HERUS & \multirow{2}{*}{1.9 (0.2)} & \multirow{2}{*}{40.2 (4.1)} & & \multirow{2}{*}{1.9 (0.2)} & \multirow{2}{*}{38.9 (3.7)} & & \multirow{2}{*}{2.0} & \multirow{2}{*}{36.6 (4.6)} \\\vspace{10pt}
                      &  (14 sources)   & &  & & &  &  &  \\
\multirow{2}{*}{Ellipticals} & Atlas$^{3 \rm D}$ & \multirow{2}{*}{1.6 (0.2)} & \multirow{2}{*}{28.9 (4.4)} & & \multirow{2}{*}{1.6 (0.2)} & \multirow{2}{*}{28.7 (4.7)} & & \multirow{2}{*}{2.0} & \multirow{2}{*}{25.3 (3.3)} \\\vspace{5pt}
                      &  (4 sources)  & &  & &  &  &  &  \\
\hline
\end{tabular}%
\end{threeparttable}
\end{table*}

For the remainder of this paper, we exclude our model with a single modified black body curve since it is prone to biases arising from the presence of a warmer dust contribution in some of our samples. Furthermore, since that the typical \BetaCold\ and \TdustCold\ of galaxies are generally difficult to estimate due to a degeneracy observed between these two parameters \citep[e.g.][]{Shetty2009, Juvela2012, Lamperti2019}, we kept our model with two modified black body curves with $\beta$ unconstrained, as well as that with $\beta$ fixed to a value of 2.

To calculate the cool dust masses for each object that could be fit, we directly measured $B_{\nu}(160,T)$ and $S_{160}$ from the fits of the cool dust component, and then used Eq.\,\ref{eq:dustMass}. These values of \Mdust\ measured from the SED fits (using $\kappa_{160}=1.038~{\rm m^2~kg^{-1}}$), and for each of our models with two modified black body curves are listed in Tables in the online material. The uncertainties on these values of \Mdust\ were measured by propagating through Eq.\,\ref{eq:dustMass} the uncertainties on each of the best fitting parameters, in turn estimated from the posterior distributions, as explained in \S\,\ref{subsubsec:IRSEDfit}.

\subsection{The dust masses from flux ratios}
\label{subsec:MdustRatio}

Because dust masses for the majority of our galaxies could not be measured using detailed SED fits due to a paucity of data, especially at the longer FIR wavelengths, we developed a method described in \S\,\ref{subsubsec:calcDustMass} to measure \Mdust\ which requires fewer FIR photometric measurements. The dust masses estimated in this way were then compared against literature values, as described in \S\,\ref{subsubsec:fitCompPast}.

\subsubsection{Method}
\label{subsubsec:calcDustMass}

The two parameters that need to be estimated to calculate \Mdust\ are the temperature and the $\beta$ index of the cold dust, using the 100 and 160~\mum\ fluxes only, since they are available for most of our galaxies (see \S\,\ref{sec:sampdat}). To do this, we used a similar approach to that presented in \cite{Tadhunter2014}. In the latter, a series of black body curves were constructed based on a \Tdust--$\beta$ grid. For objects that were detected at 100, 160, and 250~\mum, a typical $\beta~=~1.2$ was determined by comparing the observed 100/160~\mum\ and 160/250~\mum\ flux ratios to those predicted by the grid of black body curves. Finally, for all of the sources (i.e. not only those detected at 250~\mum), a new series of black body curves was generated with fixed $\beta~=~1.2$, and \Tdust\ was chosen to best match the observed 100/160~\mum\ flux ratio of each object.

For this work, we benefited from our SED fits to estimate the $\beta$ indices, instead of relying on the 100/160~\mum\ and 160/250~\mum\ flux ratios, which was the first step in \cite{Tadhunter2014}. We first adopted the mean $\beta$ index of each sample, as reported in Table\,\ref{tab:resFit} for our model with \BetaCold\ unconstrained (i.e. \BetaCold\ listed under the ``Two Comp.'' model in Table\,\ref{tab:resFit} for each sample). For the HRS sample, we adopted the mean $\beta$ index of the Atlas$^{\rm 3D}$ sample, since no galaxies could be selected for detailed SED fits (see \S\,\ref{hrs}). We also estimated dust temperatures and masses separately assuming a fixed \BetaCold~=~2 for all the samples. This allows us to gauge the effect of using different values of the $\beta$ index on our results.

The dust temperature \Tdust\ was then calculated by minimising models of black body curves with varying \Tdust\ and fixed $\beta$ indices against the observed 100/160~\mum\ flux ratios, as in \cite{Tadhunter2014}. Two sets of temperatures were derived, depending on whether \BetaCold\ was fixed to the mean of the sample as measured from the SED fits, or to a value of 2. In each case, the observed fluxes at 100~\mum\ were used to estimate the overall normalisation. The uncertainties on \Tdust\ and the normalisations were estimated by propagating the uncertainties on the fluxes, as well as on the $\beta$ indices when not fixed to a value of 2.

For galaxies that were detected only at 100 {\it or} 160~\mum\ (see Table\,\ref{tab:sample}), we had no constraints on the 100/160~\mum\ flux ratios, and the dust temperature could not be calculated as above. Instead, we adopted the mean temperatures and uncertainties for the relevant sample as found for sources which were detected at both 100 and 160~\mum, and therefore for which we could calculate the temperatures from the 100/160~\mum\ flux ratios and their uncertainties. These corresponded to \Tdust~=~34.9$\pm0.4$, 31.0$\pm0.8$, 30.4$\pm0.3$, 31.6$\pm0.3$, 32.6$\pm0.2$, and 24.3$\pm0.7$~K, for the 2Jy, 3CR, PGQs, Type-II QSOs, ULIRGs, and ellipticals (i.e. HRS and Atlas$^{\rm 3D}$ samples), respectively, when adopting the mean $\beta$ index for each sample, and \Tdust~=~30.5$\pm0.3$, 28.8$\pm0.7$, 27.0$\pm0.2$, 27.2$\pm0.2$, 31.7$\pm0.2$, and 22.3$\pm0.4$~K, respectively, when adopting a fixed $\beta~=~2$. We note that these mean temperatures are within $\sim$2--5~K of the mean \Tdust\ values found per population in our detailed SED fits (see Table\,\ref{tab:resFit}). The normalisation of the inferred black body curve and its uncertainty was then calculated based on whichever of the 100 or 160~\mum\ fluxes was detected, and by propagating the uncertainties on the fluxes and the mean parameters.

For sources with upper limits only (see Table\,\ref{tab:sample}), the upper limit at 100~\mum\ was used, along with the mean temperature for the relevant sample, and \Mdust\ was treated as an upper limit. We recall that for 20~per~cent of radio AGNs (in fact all of the upper limits in the 2Jy sample), the fluxes were treated as upper limits to account for potential non-thermal contamination, instead of true non-detections (see \S\,\ref{sec:sampdat}).

As for the SED fits (see \S\,\ref{subsubsec:fitresults_new}), we measured $B_{\nu}(160,T)$ and $S_{160}$ from the black-body curves inferred from the 100 and/or 160~\mum\ fluxes and their ratio alone. The values of \Mdust\ measured from the flux ratios (using $\kappa_{160}=1.038~{\rm m^2~kg^{-1}}$), assuming \BetaCold\ fixed to the mean of the sample, or \BetaCold\ fixed to a value of 2, are listed in Tables in the online material. The uncertainties on the values of \Mdust\ were estimated by propagating through Eq.\,\ref{eq:dustMass} the uncertainties found on each of the parameters.

By comparing the dust masses measured from detailed, two-component SED fits with unconstrained $\beta$ indices against those measured based on the 100/160~\mum\ flux ratios, where $\beta$ was fixed to the mean of the sample, we found that the flux ratio method is accurate to within a factor of two-to-five. By adopting \BetaCold~=~2 for the detailed SED fits as well as for the flux ratios, the agreement between the two methods at calculating \Mdust\ is reduced to within a factor of 1.2-to-2.5. Moreover, the dust masses for the Type-II QSOs, measured using the flux ratios with \BetaCold\ fixed to the mean value (i.e. 1.4; see Table\,\ref{tab:resFit}), are systematically higher by a factor of 1.2-to-3, compared to when measured using the flux ratios with \BetaCold~=~2 fixed. We used the Type-II QSOs for the latter comparison since they display the largest difference between the mean \BetaCold\ and the value of 2, therefore gauging the largest effect of the $\beta$ indices when using the flux ratio method to calculate values of \Mdust.

We emphasise that, although there might be systematic uncertainties in the absolute dust masses by up to a factor of five depending on the method (i.e. detailed fits versus ratio method) or $\beta$ index (i.e. \BetaCold~=~2 versus \BetaCold\ set to the mean for the sample) assumed, this will not affect the comparisons we make for our results, since we adopt a uniform approach -- based on the 100/160~\mum\ ratio method with \BetaCold~=~2 -- for all our samples.

\subsubsection{Comparison with literature values}
\label{subsubsec:fitCompPast}

\begin{figure}
\begin{center}
\includegraphics[width = 0.47\textwidth]{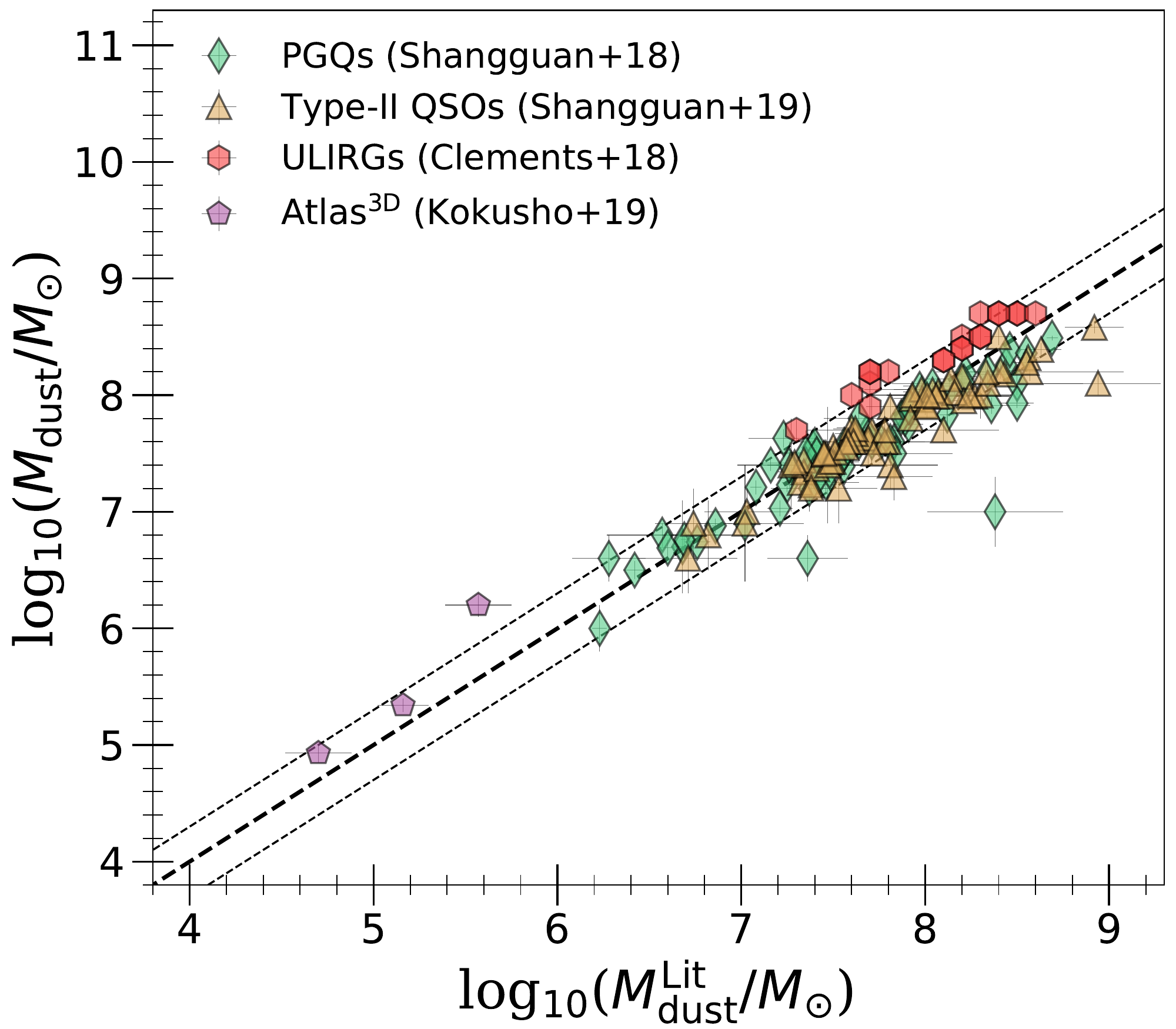}
\caption{Comparison between our dust masses inferred from the 100/160~\mum\ flux ratios (ordinate), and $M_{\rm dust}^{\rm Lit}$, found in the literature, for the PGQs \protect\citep{Shangguan2018}, Type-II QSOs \protect\citep{Shangguan2019}, ULIRGs \protect\citep{Clements2018}, and the Atlas$^{\rm 3D}$ \protect\citep{Kokusho2019} samples (see keys). In each case \BetaCold\ was chosen to best match the adopted method of the corresponding archival studies (see text). The thick and the thin dashed lines show unity, and a factor of two deviation form the latter, respectively. \label{fig:compMassPast}}
\end{center}
\end{figure}

We collected archival values of \Mdust\ to compare against those measured from the flux ratios and presented in this work. For this, we only compared objects detected at both 100 {\it and} 160~\mum. For the Atlas$^{\rm 3D}$ sample, we used the \Mdust\ estimates reported in \cite{Kokusho2019}, measured from IR SED fits assuming $\beta~=~2$. For this reason, we compared against our values of \Mdust\ calculated assuming \BetaCold~=~2. For the PGQs and the Type-II QSOs, we used the \Mdust\ values reported in \cite{Shangguan2018} and \cite{Shangguan2019}, respectively, where the dust masses were measured from SED fits using the full dust emission model of \cite{Draine2007}, after removing the AGN contributions. Because \cite{Draine2007} showed that at $\lambda~<~500$~\mum\ their models were equivalent to a modified black-body curve with $\beta~=~2$, we used for comparison our values of \Mdust\ with \BetaCold~=~2. Finally, values of \Mdust\ for ULIRGs were taken from \cite{Clements2018}, calculated via IR SED fits with a single black-body curve, and $\beta$ set as a free parameter. Therefore, we compared the latter against our \Mdust\ estimated based on the flux ratio method with \BetaCold\ set to the mean value returned by our detailed SED fits when $\beta$ was free to vary (see Table\,\ref{tab:resFit}). We stress, however, that this value of \BetaCold$\sim$~1.9 is close to the value of 2, such as choosing to compare against values of \Mdust\ calculated using the flux ratio method with \BetaCold~=~2 would not impact significantly the comparison.

We show in Fig.\,\ref{fig:compMassPast} that our values of \Mdust\ (i.e. $\sim10^{4\--9}~$\Msun) generally agree with literature values to within a factor of two, regardless of the method used. There are, however, few outliers, but which still agree to within a factor of five. The outliers in the AGN samples (i.e. mainly two PGQs; see Fig.\,\ref{fig:compMassPast}) can be explained by a larger contribution of AGN IR emission at 100~\mum\ affecting their 100/160~\mum\ flux ratios, as found when fitting their SEDs to measure SFRs (see \S\,\ref{sec:SFRcalc} for the SFRs).

We further note that the values of \Mdust\ for ULIRGs reported in \cite{Clements2018} appear systematically lower by a factor of $\sim$1.5-to-3, when compared to those measured in this work. In \cite{Clements2018}, $\beta$ was found to be 1.7, instead of $\sim$1.9 here, explaining the systematic differences in \Mdust\ for ULIRGs. The lower $\beta$ indices found in \cite{Clements2018} are likely due to the use of a single modified black-body curve, instead of our two-component approach which removes the contribution from the hotter dust (see \S\,\ref{subsubsec:IRSEDfit}). However, we stress that by using a single black-body curve, although our average $\beta$ index is slightly lower than when using two black-body curves, it remains higher than that found in \citeauthor{Clements2018} (\citeyear{Clements2018}, see Table\,\ref{tab:resFit}).

\begin{figure}
\begin{center}
\includegraphics[width = 0.47\textwidth]{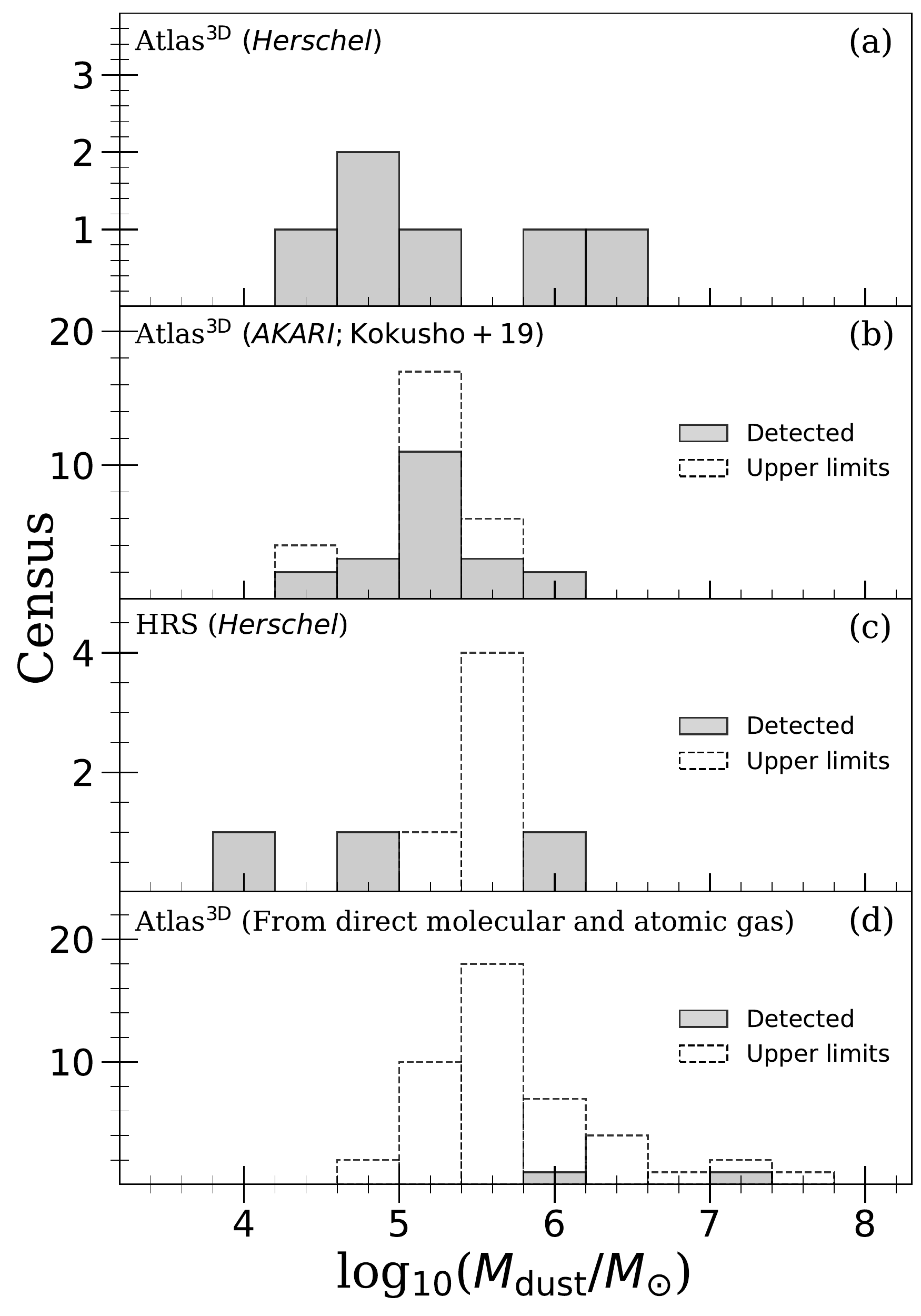}
\caption{Histogram of \Mdust\ for our samples of non-AGN elliptical galaxies. {\it Panel (a)}: the Atlas$^{\rm 3D}$ sample observed with {\it Herschel}. {\it Panel (b)}: the Atlas$^{\rm 3D}$ sample observed by AKARI \protect\citep{Kokusho2019}. {\it Panel (c)}: the HRS sample observed with {\it Herschel}. {\it Panel (d)}: the Atlas$^{\rm 3D}$ sample with \Mdust\ converted from direct measurements of molecular \protect\citep[from][]{Young2011} and atomic \protect\citep[from][]{Serra2012} gas masses. We used a factor of 140 to convert direct measurements of \Mgas\ into \Mdust\ (see text). Each bin is 0.4~dex wide, and the filled bars show the number of objects with measured \Mdust, while the dashed, open bars show the number of any additional objects with upper limits found in each bin. \label{fig:addKok}}
\end{center}
\end{figure}

Benefiting from the general agreement between the values of \Mdust\ calculated via the 100/160~\mum\ flux ratios using {\it Herschel} and those presented in the literature for the Atlas$^{\rm 3D}$ sample calculated using AKARI \citep{Kokusho2019}, we expanded our sample of elliptical galaxies using the 32 objects, including upper limits, with dust masses listed in \cite{Kokusho2019} and that were not observed by {\it Herschel} (see also \S\,\ref{atlas3d}). We show in Fig.\,\ref{fig:addKok}, panel (a), (b), and (c), that the histogram of \Mdust\ for these extra sources taken from \cite{Kokusho2019} is fully consistent with those in our {\it Herschel} samples. Therefore, including these \Mdust\ should not bias our results. 

In panel (d) of Fig.\,\ref{fig:addKok}, we show the distribution of \Mdust\ for the 45 elliptical galaxies (excluding S0 morphologies) with direct measurements of molecular and atomic gas masses (taken from \citealt{Young2011} and \citealt{Serra2012}, respectively). A typical gas-to-dust ratio of 140 was used to convert gas masses (\Mgas) into dust masses \citep[e.g.][]{Draine2007, Parkin2012}. We stress that the gas-to-dust ratio is highly uncertain \citep[e.g.][]{Kokusho2019}, and our values of \Mgas\ are only to be used as a guide. We find that the distribution of \Mdust, converted from direct measurements of \Mgas, is consistent with that measured from IR observations, although most of the former is constrained by upper limits only.

Since that the mean stellar masses of elliptical galaxies in the Atlas$^{\rm 3D}$ sample is lower than in the radio galaxy hosts (see \S\,\ref{sec:Mstarcalc} and \S\,\ref{subsec:rejuv}), there might be a concern that the comparison is not fair if the dust mass increases with stellar mass. However, there is no evidence for an increase in dust and cool ISM masses with stellar mass for elliptical galaxies \citep[e.g.][]{Young2011, Kokusho2019, Davis2019}.

\section{Measuring star formation rates}
\label{sec:SFRcalc}

In addition to the cool ISM content, we aim to compare the star-forming properties of powerful radio AGNs against our comparison samples. The SFRs of our populations of AGNs and galaxies were obtained using {\sc iragnsep}\footnote{{\sc iragnsep} is freely available at \url{https://pypi.org/project/iragnsep/}. Version 7.3.2 has been used in this work.}, which decomposes the IR SEDs of galaxies into an AGN and a galaxy contribution, therefore returning SFRs free of AGN contamination \citep[see][for details on {\sc iragnsep}]{Bernhard2021}. However, we first modified {\sc iragnsep} so that it could fit IR SEDs with FIR fluxes (i.e. $\lambda~>~100$~\mum) constrained by upper limits alone. This was useful for objects which were potentially dominated by non-thermal emission, for which we treated the fluxes as upper limits (see \S\,\ref{sec:sampdat}). For these, returned SFRs were also regarded as upper limits. In addition, we added the possibility to fit SEDs with no MIR data (i.e. $\lambda~<~70$~\mum), for which the AGN templates were not included, since they were impossible to constrain without MIR data. This was useful for objects without reliable {\it WISE} or {\it Spitzer}--MIPS fluxes. These SFRs were also regarded as upper limits, since no AGN contributions could be estimated.

For the sample of ULIRGs we set the silicate absorption parameter of {\sc iragnsep} between $S_{9.7}~=~3\--5$, as most ULIRGs show evidence of strong silicate absorption at 9.7~\mum\ \citep[e.g.][]{Rieke2009}. Assuming the optical-to-IR extinction curve of \cite{Draine2007}, these values of $S_{9.7}$ translate to $A_{\rm V}~\sim~43\--72~$mag, which are consistent with the typical values measured in star-bursting galaxies \citep[e.g.][]{Genzel1998,Siebenmorgen2007}. We stress that ignoring extinction did not change the values of SFRs significantly, but the quality of the fits was generally better once extinction was accounted for. We did not need to do this for our samples of AGNs, since the AGN emission in the MIR often dilutes the strong silicate absorption.

For each IR SED, {\sc iragnsep} fit a possible combination of 21 models (i.e. 7 different templates for galaxy emission and 2 templates for AGNs), 14 of which contain a template for the IR emission of AGNs. Each of these 21 fits are weighted using the Akaike Information Criterion \cite[e.g.][; AIC]{Akaike1973,Akaike1994}, which allows the comparison of models with a different number of degrees of freedom (i.e. those including an AGN contribution against those that do not). The best model has the highest weight \citep[see][for more details]{Bernhard2021}. To account for the fact that there is no true model, our SFRs were calculated using a weighted sum of all of the 21 possible fits, the weights of which corresponded to the Akaike weights. To estimate realistic uncertainties on the SFRs, in addition to those returned for each of the 21 fits (weighted by their AIC), we included the standard deviation of all the of 21 possible SFRs returned by the fits and weighted by their AIC. The SFRs and upper limits returned by {\sc iragnsep} for individual objects are listed in Tables available in the online material.

\section{Measuring stellar masses}
\label{sec:Mstarcalc}

To place our samples of AGNs and galaxies in the context of the main sequence of star-forming galaxies (MS), we further require stellar masses (\Mstar). For our samples of radio-loud AGNs and non-AGN elliptical galaxies, reliable values of \Mstar\ can be estimated from converting the $K_S$-band luminosities of the Two Micron All-Sky Survey (2MASS; \citealt{Skrutskie2006}) using the colour-dependent mass-to-light ratios of \cite{Bell2003}. We adopted a $B-V$ colour of 0.95, which is typical of local elliptical galaxies \citep[e.g.][]{Smith1989}, and assumed a \cite{Chabrier2003} initial mass function.

For 30 (65~per~cent) of the radio AGNs in the 2Jy sample, we used the $K$-band magnitudes\footnote{We note that, strictly speaking, the $K_{S}$-band has been used to calibrate the mass-to-light ratio in \cite{Bell2003}. When using the $K$-band magnitudes instead, we did not apply any corrections since these were found negligible compared to the typical uncertainties found for \Mstar.} provided in Table~4 of \cite{Inskip2010}, where the contribution from point sources (i.e. AGN) have been removed. This was done by decomposing images from the ESO New Technology Telescope (NTT), the United Kingdom Infra-Red Telescope (UKIRT), and the Very Large Telescope (VLT) facilities (see \citealt{Inskip2010} for details on the observations and method). For one object (i.e. PKS~0039-44), we used the $K$-band magnitude provided in Table~3 of \cite{Inskip2010}, as it was too faint to model and remove the point source. The latter paper also suggests that the $K$-band magnitude of PKS~0039-44 was not contaminated by AGN emission, and reflects the stellar emission of the host galaxy. An extra five objects in the 2Jy sample had archival Visible and Infrared Survey Telescope for Astronomy (VISTA) $K$-band magnitudes, and the redshift of one 2Jy source (i.e. PKS~0117-15) was such that the \WISE\ magnitude at 3.5~\mum\ could be converted to a $K_{S}$-band magnitude.

For the remaining sources in the 2Jy sample, as well as for the 3CR, HRS and Atlas$^{\rm 3D}$ objects, we collected archival 2MASS $K_{S}$-band magnitudes from the IRSA database. Due to the local nature of our samples, most of our sources will be spatially extended. Therefore, we primarily used extended estimates of the $K_{S}$-band magnitudes \citep{Jarrett2000}. These were available for five of the remaining 2Jy objects, 26 (58~per~cent) of the 3CR sources, and all of the HRS and Atlas$^{\rm 3D}$ galaxies. For the rest of the 2Jy and 3CR sources (i.e. two and 16 objects, respectively), we used the 2MASS point source estimate of the $K_{S}$-band magnitudes \citep{Skrutskie2006}. We estimated and corrected for the potential missed extended flux by fitting a linear relationship between the extended and the point source magnitudes, calibrated using galaxies that had both measurements available \citep[see][]{Pierce2021}.

While the majority of our $K$-band magnitudes for the 2Jy sample were corrected for potential AGN contributions, it is possible that those that were not, as well as those in the 3CR sample, suffer significant contamination by AGN emission, therefore biasing measurements of \Mstar. We found six and 13 sources in the 2Jy and 3CR samples, respectively, that were not corrected for AGN contributions, and that were also previously reported with Type-I AGN emission. The \Mstar\ values for these objects were regarded as upper limits. Finally, all of our $K$-band magnitudes have been corrected for interstellar extinction, and K-corrected using the prescription of \cite{Bell2003}, prior to calculating \Mstar. The values of \Mstar\ and upper limits are listed in Tables available in the online material.

For the PGQs, we used the \Mstar\ values provided in \cite{Zhang2016}. These were calculated by employing the same M/L method of \cite{Bell2003}, adapted for disk galaxies, when necessary, and after decomposing high-resolution optical-to-near IR images (see \S\,3 in \citealt{Zhang2016} for more details). We found direct measurements of \Mstar\ for 60~per~cent of our full sample of PGQs. For the full sample of Type-II QSOs, we used the \Mstar\ values provided in \cite{Shangguan2019}, derived from $J$-band photometry, and using the M/L ratio of \cite{Bell2001}, constrained by a $B~-~I$ colour typical of obscured QSOs (see \S\,3.1 in \citealt{Shangguan2019} for more details). Finally, we found stellar masses for 10 ULIRGs in \cite{Zaurin2010}, based on spectral synthesis modelling, and of one ULIRG in the SDSS database.

\section{Results}
\label{sec:results}

In this section, we compare the dust masses (\S\,\ref{subsec:results_dust}) and star-forming properties (\S\,\ref{subsec:results_sfr}) of our samples of powerful radio AGNs to those of our comparison samples to investigate possible differences in triggering mechanisms.

\subsection{The dust content of radio AGNs}
\label{subsec:results_dust}

\begin{figure}
\begin{center}
\includegraphics[width = 0.46\textwidth]{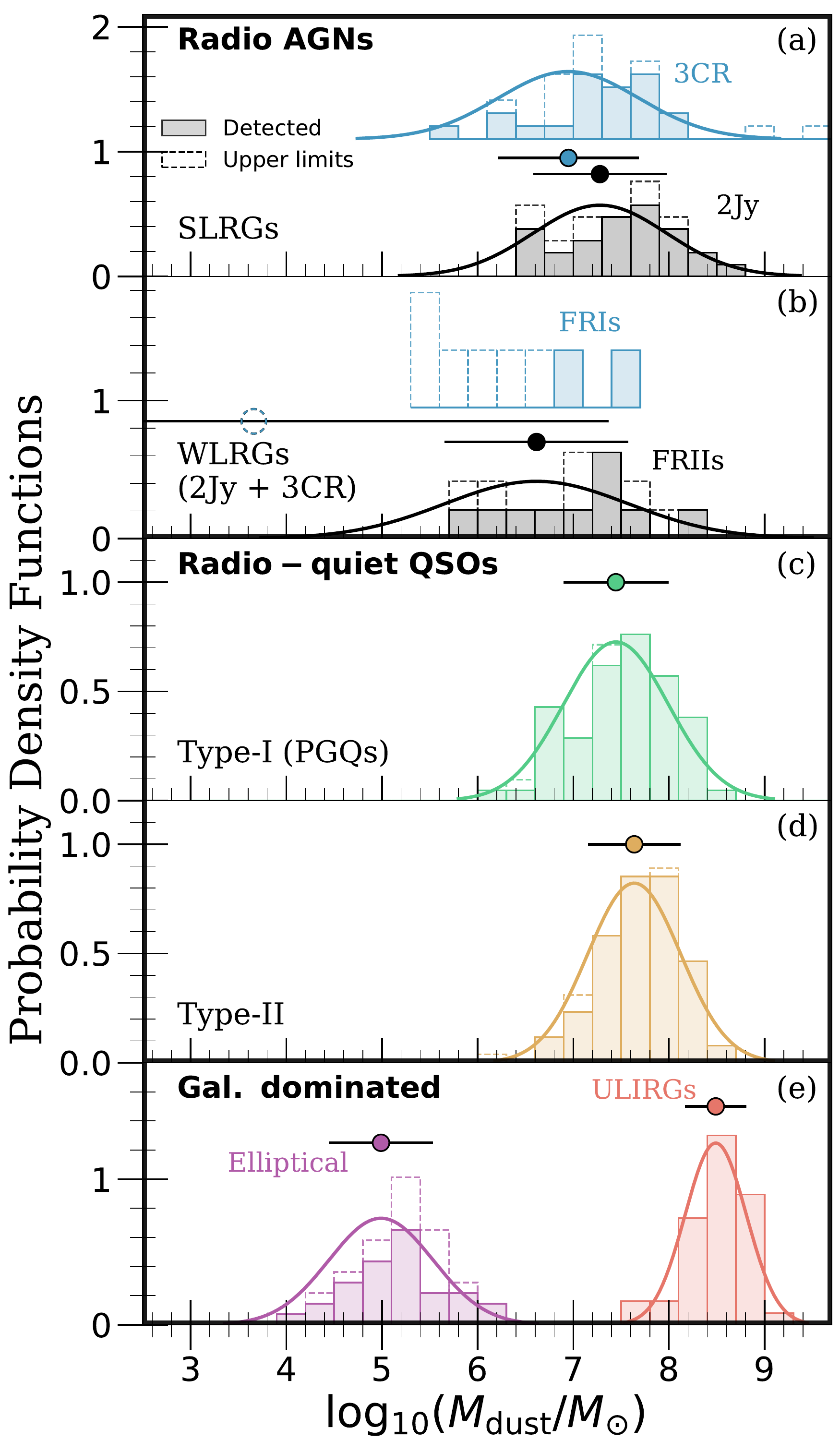}
\caption{The distributions of \Mdust\ for each of our populations and sub-samples of galaxies. Panels (a) and (b) are for radio AGNs, split in terms of SLRGs -- panel (a) -- and WLRGs -- panel (b). We further sub-divided the SLRGs into the 2Jy and 3CR samples, and the WLRGs (combining the 2Jy and 3CR samples) into FRIs and FRIIs (these further sub-divisions are shown by offsetting one of the histograms in panels (a) and (b)). Panels (c) and (d) are for radio-quiet QSOs split in terms of Type-I -- panel (c) -- and Type-II -- panel (d) -- AGNs. Panel (e) shows the galaxy dominated samples, including the non-AGN classical elliptical galaxies (purple histogram), and the ULIRGs (orange histogram). The histograms show the observed PDFs of \Mdust, where filled bars correspond to the contribution from detected sources, and empty dashed bars from any additional upper limits. The smooth log-normal distributions show the best model PDFs, which account for upper limits. The best fit parameters are shown with colour-filled circles (medians $\mu$) and error bars (uncertainties $\sigma_{\rm d}$) above or below each of their corresponding PDFs. For WLRG/FRIs we do not show the fitted PDF, and we show the average parameters with an open dashed circles since they are mostly constrained by upper limits. \label{fig:massCompFull}}
\end{center}
\end{figure}

We show in Fig.\,\ref{fig:massCompFull} the histograms of \Mdust, split in terms of samples, and where each histogram has been normalised to show the probability density function (PDF). The median and typical range of \Mdust\ for each population cannot be directly derived from the PDFs due to the presence of upper limits. Instead, we modelled the observed PDFs, assuming that the distributions of \Mdust\ are log-normal with parameters $\mu$ (the median \Mdust) and $\sigma_{\rm d}$\footnote{We used the subscript ``d'' for the standard deviations of the distributions to avoid confusion with $\sigma$, reserved to indicate the level of significance.} (the standard deviation of the distribution). We used MLE to optimise $\mu$ and $\sigma_{\rm d}$ against the observed PDFs, including upper limits on \Mdust\, and {\sc emcee} to explore the parameter space (see also \S\,\ref{subsubsec:IRSEDfit}). The $\mu$ and $\sigma_{\rm d}$ of each sample, as well as their uncertainties, measured from their posterior distributions, are listed in Table\,\ref{tab:mdustParms}. These parameters and the best fit PDFs are also shown in Fig.\,\ref{fig:massCompFull}.

\begin{table*}
\begin{threeparttable}
\caption{List of the median values ($\mu$) and standard deviations ($\sigma_{\rm d}$) for $M_{\rm dust}$ and SFRs returned by the fits of the observed PDFs (see \S\,\ref{subsec:results_dust} and \S\,\ref{subsec:results_sfr}). We split between populations, sub-populations, and sub-samples of galaxies. In this table, ``RL'', ``RQ'', and ``Gal. dom.'' refer to radio loud, radio quiet, and galaxy dominated, respectively. The numbers between brackets correspond to the estimated 1$\sigma$ uncertainties derived from the posterior distributions of each of the optimised parameters. \label{tab:mdustParms}}
\centering
\begin{tabular}{ccccccc}
\hline
Populations &  sub-pop.  &  sub-samp. & $\mu \left(\log_{10}(\frac{M_{\rm dust}}{M_\odot}) \right)$ & $\sigma_{\rm d}\left(\log_{10}(\frac{M_{\rm dust}}{M_\odot}) \right)$ & $\mu \left(\log_{10}\left(\frac{{\rm SFR}}{M_\odot/{\rm yr}} \right) \right)$ & $\sigma_{\rm d}\left(\log_{10}\left(\frac{{\rm SFR}}{M_\odot/{\rm yr}} \right) \right)$ \\
\hline
\multirow{4}{*}{RL AGNs}    & \multirow{2}{*}{SLRGs} & 2Jy                      & 7.30 (0.10) & 0.70 (0.10) &  0.70 (0.20) & 0.80 (0.10) \\ \vspace{5pt}
                            &                        & 3CR                      & 6.90 (0.20) & 0.70 (0.10) &  0.00 (0.30) & 1.20 (0.30) \\ 
                            & \multirow{2}{*}{WLRGs} & FRI                      & 4.00 (1.00) & 4.00 (2.00) & -1.00 (1.00) & 2.00 (1.00) \\ \vspace{10pt}
                            &                        & FRII                     & 6.60 (0.30) & 0.90 (0.30) & -0.30 (0.30) & 0.90 (0.30) \\
\multirow{2}{*}{RQ QSOs}    & \multirow{2}{*}{--/--} & Type-I                   & 7.45 (0.07) & 0.55 (0.05) &  0.30 (0.10) & 0.80 (0.09) \\ \vspace{10pt}
                            &                        & Type-II                  & 7.64 (0.05) & 0.47 (0.04) &  0.66 (0.08) & 0.70 (0.07) \\
\multirow{2}{*}{Gal. dom.}  & ULIRGs                 & \multirow{2}{*}{--/--}   & 8.49 (0.05) & 0.32 (0.04) &  2.04 (0.08) & 0.40 (0.06) \\
                            & Elliptical             &                          & 4.99 (0.09) & 0.54 (0.07) & -2.60 (0.40) & 1.00 (0.40) \\
\hline
\end{tabular}%
\end{threeparttable}
\end{table*}

The median dust mass of SLRGs in the 2Jy sample appears higher by a factor of 2.5 when compared to that of SLRGs in the 3CR sample (see Fig.\,\ref{fig:massCompFull} and Table\,\ref{tab:mdustParms}). However, this is only at the $\sim1.8\sigma$ level, as calculated from the \Mdust\ medians and their uncertainties (Table\,\ref{tab:mdustParms}), suggesting that the median \Mdust\ values of SLRGs are consistent between the 2Jy and 3CR samples. The discrepancy between the two is likely due to the effects of upper limits on the distributions, since the natures of these upper limits differ between the two samples. While those of the 3CR are mostly from non-detections, which affect the faintest sources, those of the 2Jy are from non-thermal contamination, which, if arising from beaming/orientation effects \citep[e.g.][]{Urry1995}, should affect a random sample of objects in the full distribution. Therefore, the median \Mdust\ value of SLRGs in the 3CR sample appears reduced to accommodate the larger number of upper limits close to the lower bound of the full distribution, when compared to the 2Jy sample.

We find a median dust mass of \Mdust~=~2~$\times~10^{7}~$\Msun, with dust masses covering a range of $10^{6}~\leq~$\Mdust$~\leq~10^{9}$~\Msun\ for the SLRGs in the 2Jy sample. The median value is a factor of two higher than that reported in \cite{Tadhunter2014} for the same sample. However, in the latter, a $\beta$ index of 1.2 was used instead of 2 in this work, and the difference in the median dust masses is fully consistent with that expected from such a difference in the $\beta$ indices (see \S\,\ref{subsubsec:calcDustMass}).

We further find that the median \Mdust\ value for WLRGs (combining the 2Jy and 3CR samples to overcome low statistics), is lower by a factor of $\sim$2000(3.3$\sigma$, hereafter the level of significance are indicated in brackets) and $\sim$5(2.2$\sigma$) for those associated with FRI-like and FRII-like radio jets, respectively, when compared to SLRGs in the 2Jy sample (see Fig.\,\ref{fig:massCompFull} and Table\,\ref{tab:mdustParms}). Therefore, it appears that the median \Mdust\ value of WLRGs/FRIIs is in better agreement with that of SLRGs in the 2Jy sample, when compared to WLRGs/FRIs. However, we note that these are based on a large number of upper limits (i.e. 60 and 50~per~cent upper limits for the WLRG/FRIs and WLRG/FRIIs, respectively) increasing the statistical uncertainties on their median \Mdust. 

To further test the differences on the average values of \Mdust\ between SLRGs and WLRGs, we focused on the 2Jy sample, since it is not affected by upper limits due to non-detections (see \S\,\ref{2Jy}). To do this, we first removed the SLRGs that are potentially contaminated by non-thermal emission (i.e. removing the upper limits on \Mdust), and were left with 27 objects out of the 35 SLRGs in the 2Jy sample. Their mean \Mdust\ can be directly calculated, since no upper limits are left, and we found $\log_{10}\left(M_{\rm dust}/M_\odot\right)~=~7.5\pm0.1$, which is in agreement with the median value of their full \Mdust\ distribution, including upper limits (see Table\,\ref{tab:mdustParms}). We then also calculated the direct mean dust masses (i.e. not using the fits of their PDFs) of WLRG/FRIs and WLRG/FRIIs, including those contaminated by non-thermal emission. Since any non-thermal contamination will boost the FIR fluxes and, therefore, the calculated dust masses, these means are likely to represent upper limits on the true mean dust masses. Out of the 11 WLRGs in the 2Jy sample, we have 6 FRIs and 5 FRIIs, of which 5 and 1 objects are potentially contaminated by non-thermal emission, respectively. The upper limits on their mean \Mdust\ were found to be $\log_{10}\left(M_{\rm dust}/M_\odot\right)~=~6.0\pm0.5$ and $\log_{10}\left(M_{\rm dust}/M_\odot\right)~=~7.1\pm0.8$ for the WLRG/FRIs and WLRG/FRIIs, respectively. Finally, we compared the mean dust mass of SLRGs (that obtained after removing the upper limits) to that of WLRGs, therefore calculating a {\it lower limit} on the differences, which constitutes a conservative approach. By doing this, we find that the mean dust masses of WLRG/FRIs and WLRG/FRIIs are lower by factors of {\it at least} $\sim$30(3$\sigma$) and $\sim$3($<$1$\sigma$) when compared to that of the SLRGs in the 2Jy sample. Therefore, at least in the 2Jy sample, and in agreement with the results of the PDF fits, it appears that the mean dust mass of WLRG/FRIs is lower when compared to that of SLRGs. In contrast, we find no clear differences in the mean dust masses of WLRG/FRIIs and SLRGs in the 2Jy sample. 

Furthermore, we do not find any significant differences between the median \Mdust\ of SLRGs in the 2Jy sample, as measured from the fits of their PDFs, and those of Type-I QSOs (difference at $\sim1.6\sigma$ level). By contrast, we find that the median \Mdust\ of SLRGs in the 2Jy sample is {\it significantly} lower by a factor of $\sim$2.2(3$\sigma$) when compared to Type-II QSOs. However, we note that, when considering the sub-sample of SLRGs in the 2Jy sample with \Loiii~$>~4.2~\times~10^{34}~$W, which is the lowest \Loiii\ luminosity in our Type-II QSO sample, the difference is less significant (difference at 2.1$\sigma$ level). Our median values of \Mdust\ for Type-I and Type-II QSOs are also in excellent agreement with those reported in \cite{Shangguan2019}, and no significant differences are found between the median \Mdust\ of these two classes (difference at $\sim2.2\sigma$ level).

Finally, using the medians from the fits to the PDFs, we find that the median \Mdust\ of SLRGs in the 2Jy sample and that of WLRG/FRIIs in the 2Jy and 3CR samples are enhanced by factors of $\sim$200(17$\sigma$) and $\sim$40(5$\sigma$), respectively, when compared to those of non-AGN classical elliptical galaxies. In contrast, the median \Mdust\ of WLRG/FRIs (although weakly constrained) is consistent with that of non-AGN elliptical galaxies (difference at $\sim1\sigma$ level). We note that the PDFs of the latter have been built by combining the Atlas$^{\rm 3D}$ and HRS samples with {\it Herschel} observations, as well as dust masses of elliptical galaxies observed with AKARI taken from \citeauthor{Kokusho2019} (\citeyear{Kokusho2019}; see \S\,\ref{subsubsec:fitCompPast}). It is also striking that these medians are lower by a factor of $\sim$16(11$\sigma$), $\sim$31000(5$\sigma$), and $\sim$80(6$\sigma$), for the SLRGs, WLRG/FRIs, and WLRG/FRIIs, respectively, when compared to that of ULIRGs. These are consistent with early results reported in \cite{Tadhunter2014} for SLRGs only. However, we note that the model PDFs do show some overlap between these populations of galaxies (see Fig.\,\ref{fig:massCompFull}).

\subsection{The star formation rates of radio AGNs}
\label{subsec:results_sfr}

\begin{figure}
\begin{center}
\includegraphics[width = 0.46\textwidth]{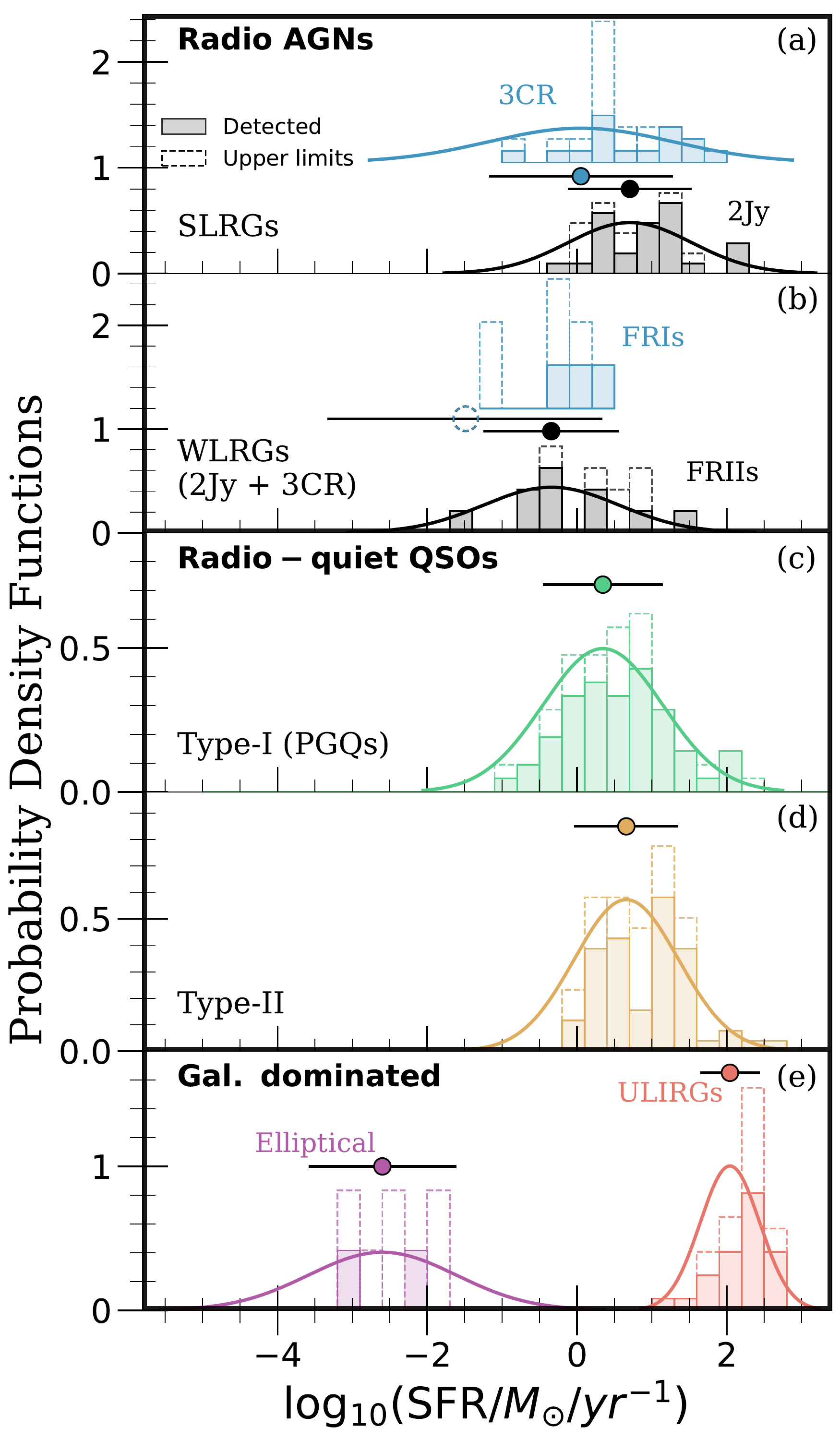}
\caption{Same as Fig\,\ref{fig:massCompFull}, but for the SFRs. \label{fig:SFRCompFull}}
\end{center}
\end{figure}

We show in Fig.\,\ref{fig:SFRCompFull} the histograms of SFRs, split in terms of galaxy populations and samples. Each histogram has been normalised to show the PDF. As for the distributions of \Mdust, we measured the median SFRs ($\mu$) and standard deviations ($\sigma_{\rm d}$) of each sample by fitting log-normal distributions to the observed histograms, including upper limits on SFRs (see \S\,\ref{subsec:results_dust}). The resulting statistics are listed in Table\,\ref{tab:mdustParms} and shown in Fig.\,\ref{fig:SFRCompFull}.

We find that the median SFRs of SLRGs in the 2Jy sample is higher by a factor of $\sim$4(1.7$\sigma$), when compared to that of the SLRGs in the 3CR sample (see Fig.\,\ref{fig:SFRCompFull} and Table\,\ref{tab:mdustParms}). The level of significance suggests that the median SFRs of SLRGs in the two samples are consistent within the uncertainties, especially considering the different effects that the upper limits might have on the two samples (see \S\,\ref{subsec:results_dust}): it is likely that the apparent differences between the two distributions can be attributed to the large number of SFR upper limits for the 3CR sample (i.e. $\sim$50~per~cent), when compared to the 2Jy sample (i.e. $\sim$30~per~cent). This acts to reduce the median SFR and increase the typical range of \Mdust\ in the 3CR sample, when compared to the 2Jy sample (see Table\,\ref{tab:mdustParms}). Adopting the SFRs of SLRGs in the 2Jy sample, since better constrained, the median SFR is $\sim$~5~\Msun~yr$^{-1}$, and the values span the range 0.3--300~\Msun~yr$^{-1}$.

We further find that the median SFR of WLRGs (combining the 2Jy and 3CR samples to overcome low statistics) is lower by factors of $\sim$50(1.7$\sigma$) and $\sim$10(2.8$\sigma$) for those associated with FRI-like and FRII-like radio jets, respectively, when compared to SLRGs in the 2Jy sample (see Fig.\,\ref{fig:SFRCompFull} and Table\,\ref{tab:mdustParms}). Because of the potential effects of upper limits, we performed a similar analysis to that presented in \S\,\ref{subsec:results_dust} for the dust masses, allowing us to derive a conservative difference between the median SFRs of SLRGs and WLRGs in the 2Jy sample. In doing this, we found that the median SFRs of WLRG/FRIs and WLRG/FRIIs in the 2Jy sample are lower by factors of {\it at least} $\sim$30(7$\sigma$) and $\sim$6(2$\sigma$), respectively, when compared to the median SFR of SLRGs in the 2Jy sample. Consistent with our results on the median dust masses of these populations, the median SFR of WLRG/FRIIs appears to be in better agreement with that of SLRGs, at least in the 2Jy sample, compared to when considering the difference between WLRG/FRIs and SLRGs.

We also find that the SFRs of the SLRGs in the 2Jy sample are fully consistent with those of Type I and Type II radio-quiet QSOs. Furthermore, there are no differences between the PDFs for SFR of Type-I and Type-II QSOs, consistent with previous work \citep[e.g.][]{Shangguan2019, Mountrichas2021}.

Finally, using the medians from the fits to the PDFs, we find that the median SFRs of SLRGs in the 2Jy sample and that of WLRG/FRIIs in the 2Jy and 3CR samples are enhanced by factors of $\sim$2000(7.4$\sigma$) and $\sim$200(4.6$\sigma$), respectively, when compared to those of non-AGN classical elliptical galaxies. In contrast, although the median SFR of WLRG/FRIs appears weakly constrained, it is lower by a factor of $\sim$40(1.5$\sigma$) when compared to non-AGN elliptical galaxies. On the other hand, these median values are lower by factors of $\sim$20(6.2$\sigma$), 1100(3$\sigma$), and $\sim$200(7.5$\sigma$), for the SLRGs, WLRG/FRIs, and WLRG/FRIIs, respectively, when compared to that of ULIRGs. This follows a similar pattern to the results reported for the dust masses of radio AGNs in \S\,\ref{subsec:results_dust}. However, it is important to add the caveat that the dust masses and SFRs are not entirely independent, since they have both been calculated using the FIR luminosities.

\section{Discussion}
\label{sec:discussion}

In this section, we explore the implications of the results for our understanding of triggering and feedback in radio AGNs. In particular, we first discuss whether the cool ISM masses found are sufficient to power QSO-like activity (\S\,\ref{subsec:QSOactivity}). Then, in \S\,\ref{subsec:discussDistrib} and \S\,\ref{subsec:AGNvsMgas} we explore the triggering mechanisms of AGNs and how they connect to the amount of gas available for both AGN activity and star formation. In \S\,\ref{subsec:AGNvsSFefficiencies} we discuss the potential impact of AGN feedback on AGN hosts. Finally, in \S\,\ref{subsec:rejuv}, we place our samples of AGNs in the broader context of the MS of galaxies, and discuss the implications in terms of their triggering mechanisms.

\subsection{Sustaining powerful QSO activity}
\label{subsec:QSOactivity}

In \S\,\ref{subsec:results_dust}, we established that the dust masses of SLRGs in the 2Jy sample, which are better constrained than those of the 3CR, have a median of \Mdust~=~2~$\times~10^{7}~$\Msun\ and span \Mdust~$\sim~10^{6\--9}$~\Msun. Assuming a typical gas-to-dust ratio of 140 (see \S\,\ref{subsubsec:fitCompPast}), we find that the median gas mass of SLRGs is \Mgas~$=~2.8~\times~10^{9}$~\Msun\ with a range of \Mgas~$\sim~10^{8\--11}$~\Msun. If we follow the arguments proposed in \cite{Tadhunter2014}, assuming QSO bolometric luminosities $L_{\rm bol}~>~10^{38}$~W, and a radiative efficiency of 10~per~cent, a mass inflow rate of $\dot{M}~>~0.2$~\Msun~${\rm yr}^{-1}$ onto the supermassive black-hole is required to sustain QSO activity. Widely varying constraints on the lifetime of QSOs suggest that QSOs are ``on'' for (i.e. duty cycle) $t_{\rm dc}~\sim~10^{6\--9}~$yr \citep[e.g.][]{Martini2004, Adelberger2005, Croom2005, Shen2009, White2012, Conroy2013}. Therefore, the total mass accreted onto the black hole during QSO episodes is $M_{\rm acc}~>~2~\times~10^{5\--8}$~\Msun. However, the gas feeding such AGN episode is likely to have originated in a gas reservoir at larger scales; this reservoir will eventually form stars in the bulge of the host galaxy. Indeed, in order to maintain the observed black hole-to-bulge mass relationship, the total mass of the gas reservoir is required to be $\sim$500 times larger than the mass of gas accreted by the black hole \cite[e.g.][]{Marconi2003}. Therefore, for the black hole to accrete $M_{\rm acc}~>~2~\times~10^{5\--8}$~\Msun\ for $t_{\rm dc}~\sim~10^{6\--9}~$yr, a total reservoir containing $M_{\rm gas}^{\rm tot}~\gtrsim~10^{8\--11}$~\Msun\ of gas is required. These estimates of $M_{\rm tot}$ are in remarkable agreement with the range of \Mgas\ estimated in this work for SLRGs.

We further found evidence for WLRGs associated with FRI-like radio jets to have a lower median dust mass when compared to that of SLRGs (see \S\,\ref{subsec:results_dust}). The values span the range \Mdust~$\sim~10^{5\--8}$~\Msun, which translate to \Mgas~$\sim~10^{7\--10}$~\Msun, respectively. Following the aforementioned argument, these overlap with the gas masses necessary to trigger powerful radiatively-efficient QSOs. Therefore, at least for some of the most gas-rich WLRGs associated with FRIs, the presence of a substantial gas reservoir is not a sufficient condition to trigger a powerful QSO. In this case, other factors such as the detailed distribution and dynamics of the cool ISM are likely to be important \citep[e.g.][]{Tadhunter2014}. In fact, as the cool ISM settles into a dynamically stable configuration post merger, the rate of gas infall to the black hole is expected to drop, leading to a lower level of nuclear activity and perhaps a WLRG AGN classification \citep[see][and references therein]{Tadhunter2011}.

\subsection{The importance of mergers for triggering powerful radio QSOs}
\label{subsec:discussDistrib}

We found that the cool ISM masses of SLRGs in the 2Jy sample are enhanced by a factor of $\sim$200(17$\sigma$), when compared to our sample of non-AGN classical elliptical galaxies (see \S\,\ref{subsec:results_dust} and Table\,\ref{tab:mdustParms}). We also recall that the latter is likely to be more FIR bright and dust-rich than typical elliptical galaxies of similar stellar mass in the local universe due to selection effects (see \S\,\ref{subsec:samp_ellip}). Therefore the quoted factor of $\sim$200 corresponds to a lower limit. Because powerful radio AGNs mostly reside in elliptical galaxies, there must be some mechanisms at work to enhance the cool ISM masses of the elliptical hosts, which in turns could be connected to the triggering of the AGN. Compelling evidence has been found in deep optical imaging for a high incidence of tidal features and double nuclei, strongly suggesting that galaxy mergers and interactions are important for their triggering \citep[e.g.][]{RamosAlmeida2011, RamosAlmeida2012, Pierce2021}. However, the fact that we found that the median \Mdust\ and SFR of SLRGs are significantly lower than those of ULIRGs, implies that for most objects the triggering mergers and interactions are likely to have been relatively minor in terms of their cool ISM contents. This is consistent with the evidence that population of massive elliptical galaxies has mainly evolved via minor mergers since \z~$\sim$~1 \citep[e.g.][]{Bundy2009, Kaviraj2009}.

Although the majority of SLRGs are unlikely to be triggered at the peaks of major gas-rich galaxy mergers, there is a significant overlap between the values of \Mdust\ (and SFRs) for SLRGs and ULIRGs (see Fig.\,\ref{fig:massCompFull} and Fig.\,\ref{fig:SFRCompFull}). To estimate the fraction of SLRGs in our samples that could be triggered by a ULIRG-like major gas-rich merger, we calculated the overlapping fraction between the PDFs of \Mdust\ for our samples of SLRGs and ULIRGs (see \S\,\ref{subsec:results_dust}, Fig.\,\ref{fig:massCompFull}, and Table\,\ref{tab:mdustParms} for the PDFs). We found fractions of 22$^{+11}_{-10}$~per~cent and 11$^{+8}_{-7}$~per~cent for the SLRGs in the 2Jy and 3CR samples, respectively. Repeating this comparison for the PDFs of SFRs led to similar summary statistics. In addition, we find that 4 out of the 5 SLRGs in the 2Jy sample (80~per~cent) with \Mdust~$>~10^{8}$~\Msun, a value consistent with the typical dust masses measured for ULIRGs, are also reported to have strong poly-aromatic hydrocarbon (PAH) emission at MIR wavelengths, which is a sign of ongoing star formation \citep{Dicken2012}.

Similarly to the SLRGs, we also found that the cool ISM properties of WLRG/FRIIs were significantly enhanced compared to those of classical elliptical galaxies, yet depleted when compared to ULIRGs (see \S\,\ref{subsec:results_dust}). In contrast, the cool ISM properties of WLRG/FRIs were not found significantly enhanced when compared to those of classical elliptical galaxies, suggesting that both populations could be consistent in terms of their cool ISM properties. These results could also imply different triggering mechanisms between SLRGs and WLRGs/FRIs, in agreement with previous work \citep[e.g.][]{Hardcastle2007, Buttiglione2009, Tadhunter2011}. This is also supported by recent evidence from deep optical imaging that mergers are less important for WLRGs compared to SLRGs \citep[e.g.][]{RamosAlmeida2011, Pierce2021}. An alternative fuelling scenario for WLRGs involves the direct accretion of the hot ISM \citep{Best2005, Allen2006, Best2006, Hardcastle2007, Buttiglione2009}, which is possible in our samples of WLRG/FRIs since we find a smaller amount of cool ISM when compared to SLRGs. Finally, the similarities found between the cool ISM properties of WLRG/FRIIs and SLRGs is consistent with the idea that the AGNs in the WLRG/FRIIs have recently switched off, and the information has not yet reached the hotspots of the radio lobes \cite[e.g.][]{Buttiglione2010, Tadhunter2012}.

\subsection{The lack of relationship between AGN power and gas mass}
\label{subsec:AGNvsMgas}

\begin{figure*}
\begin{center}
\includegraphics[width = \textwidth]{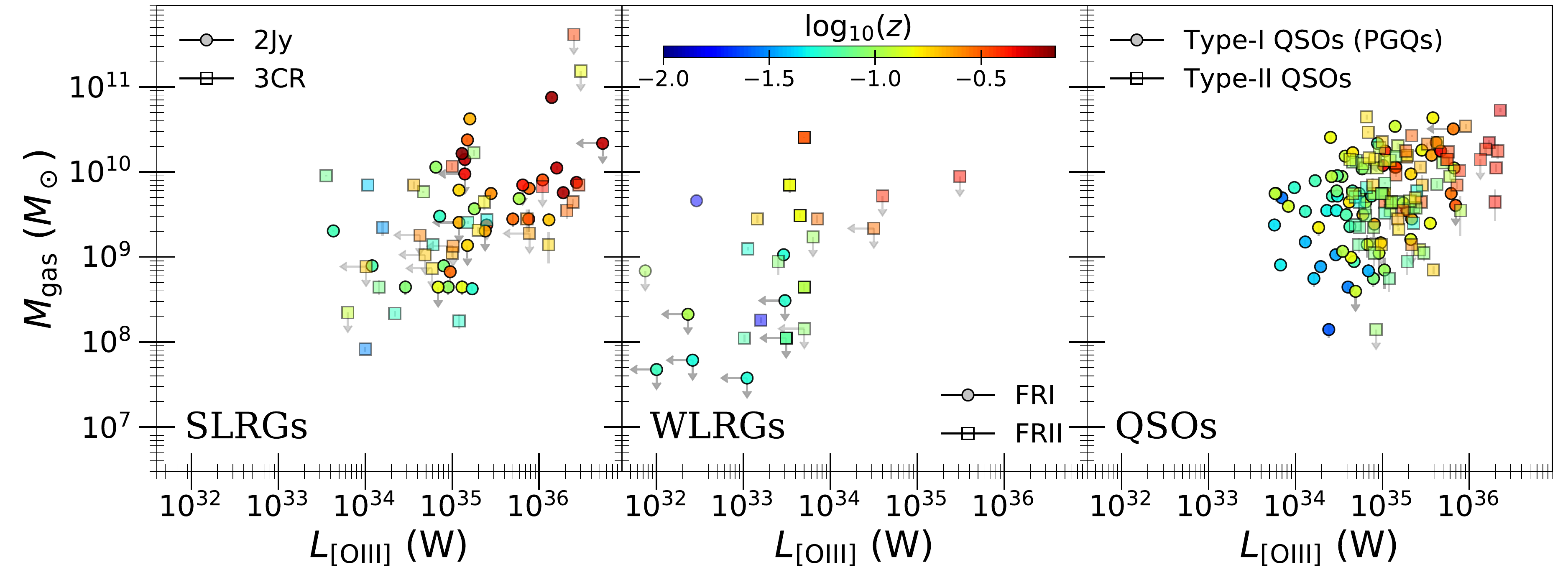}
\caption{The cool ISM masses of our samples of AGNs versus \Loiii, a proxy of AGN power. In reading order, we have the SLRGs, split in terms of the 2Jy and 3CR samples, the WLRGs, split in terms of FRI and FRII radio morphologies, and the radio-quiet QSOs, split in terms of Type-I (PGQs) and Type-II AGNs. Arrows indicate upper limits. The colour-code indicates redshift, as shown by the colour bar at the top of the central panel (0.01~$<$~\z~$<$~0.7). \label{fig:gasvsoiii}}
\end{center}
\end{figure*}

In \S\,\ref{subsec:discussDistrib} we suggested that the differences in the dust masses between SLRGs and WLRG/FRIs relate to them being triggered by different mechanisms. We now investigate whether there is a direct relationship between \Mgas\ and \Loiii\ for these populations, where \Loiii\ can be used to trace AGN bolometric luminosity \citep[e.g.][]{Heckman2005, Stern2012, Dicken2014}. Fig.\,\ref{fig:gasvsoiii} shows that there is a strong apparent relationship between \Mgas\ and \Loiii\ across several orders of magnitude in both quantities for SLRGs, WLRGs, and radio-quiet QSOs. However, once split in terms of redshift (over \z~$\sim$~0.01 to \z~$\sim$~0.7), we also find a strong relationship with redshift, suggesting that the apparent connection between \Mgas\ and \Loiii\ is fully driven by redshift (i.e. Malmquist bias). This is consistent with results from \cite{Shangguan2019}, where no relationships were found between \Mgas\ and the bolometric luminosities of the PGQs and Type-II QSOs. In an attempt to quantify this, we performed multi-linear regressions between \Mgas, \Loiii, and \z\ for our samples of SLRGs and WLRGs. We found no relationships between \Mgas\ and \Loiii\ once redshift was accounted for.

In Fig\,\ref{fig:gasvsoiii}, it also appears that a minimum gas reservoir of \Mgas~$\sim~10^{8}~$\Msun\ is required to trigger radiatively efficient AGNs, as represented by SLRGs and radio-quiet QSOs. However, the lack of a direct relationships between \Mgas\ and the power of the AGN, suggests that, although a minimum gas reservoir is likely necessary to trigger the most powerful AGNs, it is not by itself sufficient to explain the range of AGN properties: other factors, such as the gas distribution, extend to which the gas has settled into a stable dynamical configuration, and overall gas dynamics are also likely to be important (see also \S\,\ref{subsec:discussDistrib}).

\subsection{AGN power versus star-formation efficiencies: the effect of AGN feedback}
\label{subsec:AGNvsSFefficiencies}

\begin{figure*}
\begin{center}
\includegraphics[width = \textwidth]{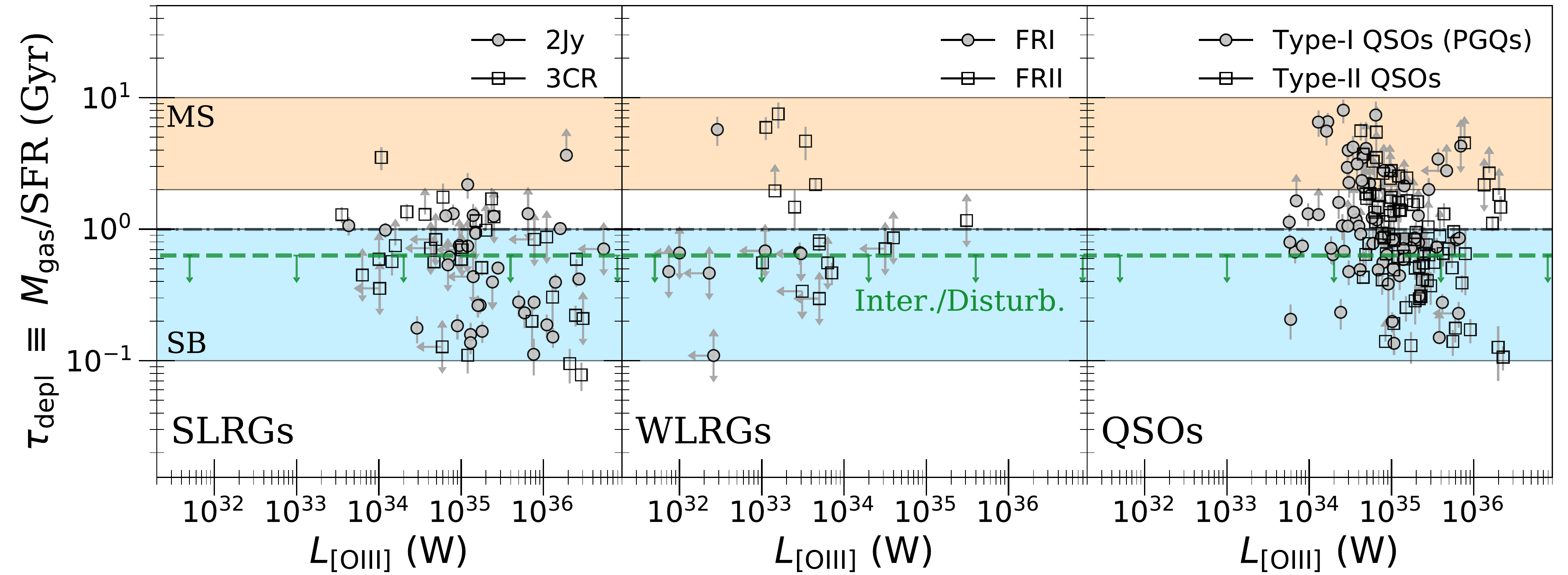}
\caption{The depletion timescales $\tau_{\rm depl}$, measured as \Mgas/SFR and expressed in Gyr, versus \Loiii\ (or AGN power) for our samples of SLRGs, WLRGs, and radio quiet QSOs (from left-to-right, respectively). We split the SLRGs into the 2Jy and 3CR samples, the WLRGs into FRIs and FRIIs, and the QSOs into Type-Is (PGQs) and Type-IIs (see keys). Arrows indicate upper and lower limits. The filled orange and blue areas indicate the range of $\tau_{\rm depl}$ typically observed in star-forming (MS) and star-bursting (SB) galaxies, respectively \protect\citep{Kennicutt1998}. The horizontal dashed green line with downward arrows show the $\tau_{\rm depl}$ below which most of interacting galaxies are expected to be found, as taken from \protect\cite{Saintonge2012}.\label{fig:efficiencies}}
\end{center}
\end{figure*}

In the most recent cosmological simulations, AGN feedback is used to regulate star formation in order to reproduce the local scaling relationships between the black hole and bulge masses, as well as the galaxy mass function \citep[e.g.][]{Schaye2015}. The net effect of such AGN feedback is a suppression of the ``in-situ'' SFRs, via the heating and/or removal of the cold gas \citep[e.g.][]{DiMatteo2005, Zubovas2012, Costa2018}. While considerable evidence now exists that AGN drive powerful, multi-phase outflows that are likely to affect the host galaxies {\it at some level} (see \citealt{Fabian2012} for a review), aside from few individual objects \citep[e.g.][]{Nesvabda2010, Lanz2016, Nesvabda2021} the direct impact that these outflows have on SFRs remains uncertain. Indeed, most statistical studies do not find any clear signs of ``in-situ'' SFR suppression as a consequence of AGN feedback \cite[e.g.][]{Maiolino1997, Stanley2015, Rosario2018, Shangguan2018, Ellison2019, Shangguan2019, Jarvis2020, Shangguan2020, Yesuf2020}.

To investigate this in our samples of AGNs, we plot in Fig.\,\ref{fig:efficiencies} the depletion timescale $\tau_{\rm depl}$, calculated using \Mgas/SFR (expressed in Gyr), against \Loiii\ for the SLRGs, WLRG/FRIs, WLRG/FRIIs, and radio-quiet QSOs. Note that potential correlations with redshift can be ignored in the case of $\tau_{\rm depl}$, because both the gas masses and SFRs would be affected in a similar way. No clear correlations are found between $\tau_{\rm depl}$, which represents the inverse of the star formation efficiency, and \Loiii, and there is a considerable scatter in $\tau_{\rm depl}$ for most \Loiii\ values.

Interestingly, we find that SLRGs mostly show shorter $\tau_{\rm depl}$ (i.e. $\lesssim1$~Gyr), when compared to WLRGs, implying vigorous SFRs for their gas masses (i.e. 5~\Msun~yr$^{-1}$ on average). Roughly half of our radio-quiet QSOs also show such short values of $\tau_{\rm depl}$, which are typically observed in star-bursting galaxies, and suggest high star-formation efficiencies where the gas is consumed rapidly \citep[e.g.][]{Kennicutt1998}. 

We also show in Fig.\,\ref{fig:efficiencies} with a dashed-green line the mean value of $\tau_{\rm depl}$ found for interacting and disturbed nearby galaxies (including major mergers) in the full sample of the CO Legacy Database for the Galex-Arecibo-SDSS Survey \citep[COLD GASS][]{Saintonge2011}, as reported in \cite{Saintonge2012}, and based on molecular gas measurements. We stress that, in the latter, although the majority of interacting systems were found with shorter values of $\tau_{\rm depl}$ (i.e. $<1$~Gyr), not {\it all} galaxies in their sample with shorter values of $\tau_{\rm depl}$ were interacting systems, and their control sample (i.e. non-interacting systems) spanned a large range of values (i.e. 0.5-to-5~Gyr), largely overlapping with those of interacting systems, but with a mean of $\sim$1~Gyr.

We note that the SLRGs in our samples display a heavily skewed distribution toward shorter $\tau_{\rm depl}$ values, consistent with those typically measured for interacting systems, suggesting that there is an excess of galaxies in that region, compared to the general population (see Fig.\,\ref{fig:efficiencies}). This is consistent with the idea that SLRGs are mainly triggered in interacting systems (see \S\,\ref{subsec:discussDistrib}). In contrast, the distribution of $\tau_{\rm depl}$ for WLRGs and radio-quiet QSOs appears randomly distributed around to the mean value of the full sample of \citeauthor{Saintonge2012} (\citeyear{Saintonge2011}; i.e. $\sim$1~Gyr). This is consistent with them being triggered in a range of situations, some of which will be connected to a merger, and consistent with the lesser role of mergers in WLRGs when compared to SLRGs.

Overall, we do not find any signs of reduced star-formation efficiencies, which is an expected outcome of AGN feedback. We stress, however, that IR-based SFRs are averaged over $\sim$100~million years. Therefore, it is possible that the effect of AGN feedback on the star-formation efficiencies is not yet apparent.

\subsection{Powerful radio AGNs: link with the rejuvenation of galaxies}
\label{subsec:rejuv}

\begin{figure}
\begin{center}
\includegraphics[width = 0.47\textwidth]{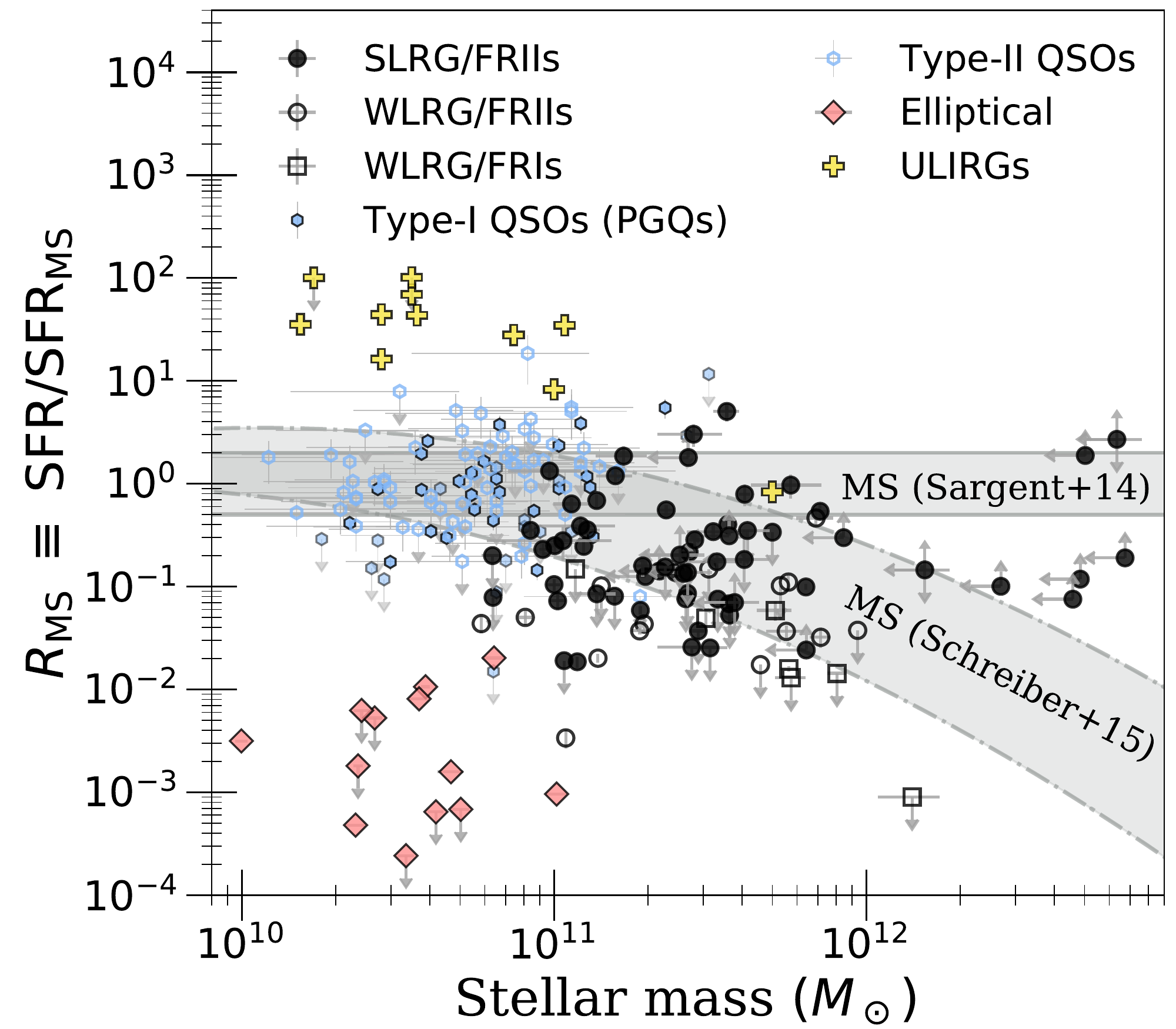}
\caption{The SFRs relative to that of the MS (\Rms) of \protect\cite{Sargent2014} versus \Mstar\ for our galaxy populations. We separated radio AGNs into SLRGs, WLRG/FRIs, and WLRG/FRIIs, as well as radio-quiet QSOs into Type-I and Type-II (see keys). Arrows indicate upper and/or lower limits on \Rms\ and \Mstar. The scatter around the MS of \protect\cite{Sargent2014} is shown with a grey band delimited by continuous lines. The curved grey bands delimited by dot-dashed lines show the deviations of the MS of \protect\cite{Schreiber2015} from that of \protect\cite{Sargent2014}, at 0.01~$<$~\z~$<$~0.7, and including the scatter.} This deviation could be due to a population of rejuvenated galaxies and coincides with the \Rms\ of our samples of powerful radio AGNs.\label{fig:SFRMS}
\end{center}
\end{figure}

To place our AGN samples in the broader context of galaxy evolution, we now compare the SFRs of our AGN hosts to those expected from the main sequence (MS) of galaxies. This is partly motivated by suggestions that AGN feedback plays an important role in the rapid quenching of galaxies, placing them below the MS \citep[e.g.][]{Smethurst2016}. To test this, we calculated \Rms~$\equiv$~SFR/SFR$_{\rm MS}$, where SFR$_{\rm MS}$ is the corresponding MS SFR at a given \Mstar\ and redshift. We use the MS of \cite{Sargent2014}, who defined a linear relationship between $\log_{10}$(SFR) and $\log_{10}$(\Mstar) for star-forming galaxies. The results are depicted in Fig.\,\ref{fig:SFRMS}.

We first note that the elliptical host galaxies of most of our powerful radio-loud AGNs have high stellar masses (\Mstar~$>~10^{11}$~\Msun), in contrast to our samples of radio-quiet QSOs which tend to have lower values of \Mstar, on average. This is in agreement with many past studies on the stellar masses of powerful radio AGNs \citep[e.g.][]{Dunlop2003, Inskip2010, Tadhunter2011}. We further find that the values of \Rms\ for powerful radio AGNs are consistent with a large range of values, from those typical of classical elliptical galaxies (i.e. \Rms~$\sim$~0.001) to those of strongly star-bursting systems (i.e. \Rms~$\sim$~4), but remain below those of most ULIRGs (i.e. \Rms~$\sim$~10\--100). Interestingly, the majority of powerful radio AGNs are located below the MS of \cite{Sargent2014}, at given \Mstar\ and redshift. This contrasts with radio-quiet QSOs, that are typically found to be consistent with the MS, in agreement with the results of \cite{Shangguan2019}. The latter authors also reported no differences in the \Rms\ values between Type-I and Type-II QSOs, as we find here.

It is notable that the SLRGs show a wide range of \Rms\ values. Moreover, while there is an overlap in the distributions, the \Rms\ values of the SLRGs are higher on average than those of WLRGs, particularly the WLRG/FRIs sources whose \Rms\ values all lie well below the MS of \citeauthor{Sargent2014} (\citeyear{Sargent2014}; see Fig.\,\ref{fig:SFRMS}). This is consistent with WLRGs having lower SFRs when compared to SLRGs, since they are hosted by galaxies with similar stellar masses at similar redshifts. It is also consistent with the longer depletion timescales found in WLRGs when compared to SLRGs (see \S\,\ref{subsec:AGNvsSFefficiencies}). Therefore, since at least some WLRGs also appear to have significant cool ISM masses (see \S\,\ref{subsec:discussDistrib}), it implies that some mechanisms are at work to reduce the efficiency of both AGN activity and star formation in WLRGs. For example, the gas might have settled to a more stable dynamical configuration in the WLRGs (see \S\,\ref{subsec:QSOactivity}). In this context, we note that there is evidence for reduced star formation efficiencies in early-type galaxies that have accreted gas through minor mergers with gas-rich galaxies \citep{Davis2015}, and for which the gas appears to have reached a relatively dynamically settled state.

One attractive explanation for our finding that radio-loud AGNs tend to fall below the MS, in contrast to radio-quiet QSOs, is that radio jets play a key role in quenching star formation, as expected from some cosmological simulations. However, this picture is inconsistent with the fact that the majority of SLRGs have relatively high star-formation efficiencies which do not show any signs of deficiency (see \S\,\ref{subsec:AGNvsSFefficiencies}). Therefore, it is more plausible that we are witnessing the late-time retriggering of galaxies, in terms of star formation and AGN activity \citep[e.g.][]{Tadhunter2014}.

Late-time re-triggering is also supported by the location of our powerful radio AGNs in the \Rms-\Mstar\ parameter space. In fact, they coincide with a specific type of massive galaxy in the local Universe that are believed to be rejuvenated (i.e. re-triggered), and for which a quenched bulge was formed early on (i.e. typically \z~$>$~2), followed by a more recent burst of star formation \citep[e.g.][]{Clemens2009, Thomas2010, Pandya2017, Chauke2019}. This has the effect of moving the otherwise quiescent systems toward the MS of galaxies. Such rejuvenation has also been used to explain the apparent curvature in the MS at high stellar masses (\Mstar~$>~10^{11}~$\Msun) that has been found in some studies \citep[e.g.][]{Schreiber2015}: as the (quiescent) galaxy bulges become more dominant at high stellar masses, any star formation due to late-time gas accretion becomes less closely tied to the total stellar mass, so the objects fall further below the MS \citep[e.g.][]{Mancini2019}. The position of the curved MS for star-forming galaxies of \cite{Schreiber2015} for 0.01$~<~$\z$~<~$0.7, relative to the MS of \cite{Sargent2014} is shown in Fig.\,\ref{fig:SFRMS} (including the scatter).\footnote{We stress that here we are comparing with the positions of the MS derived for objects pre-identified as star-forming galaxies, rather that those derived for samples that include a mixture of star-forming and quiescent, red and dead, elliptical galaxies \citep[e.g.][]{Eales2017}, which tend to fall at lower SFRs for a given stellar mass and naturally show a curve at higher stellar masses.}

Interestingly, most of our powerful radio-loud AGNs lie on the curve of the MS of \cite{Schreiber2015}, where there is significant deviation from the linear MS of \cite{Sargent2014}, perhaps due to a population of rejuvenated galaxies (see Fig\,\ref{fig:SFRMS}). In contrast, a relatively high proportion of WLRGs fall below the curve of the MS of \cite{Schreiber2015}, again emphasising their lower SFRs, which are perhaps related to triggering mechanisms that are different from those of SLRGs (e.g. direct accretion of hot gas from the X-ray haloes). Relative to the curved MS of \cite{Schreiber2015}, radio-quiet QSOs remain on the MS, since they fall within a stellar mass regime where both the linear and curved MS agree. Therefore, the triggering of the activity in these galaxies is less likely to be related to rejuvenation.

\section{Conclusion}
\label{sec:conclusion}

Taking advantage of recent IR observing campaigns undertaken for the 2Jy sample, and the availability of archival IR data (see \S\,\ref{sec:sampdat}), we investigated the triggering and feedback mechanisms of powerful radio AGNs, split in terms of SLRGs (i.e. showing strong optical emission lines, typically observed in QSOs) and WLRGs (i.e. lacking strong optical emission lines). To do this, we calculated dust masses (see \S\,\ref{sec:dustMass}), tracing the cool ISM, and SFRs (see \S\,\ref{sec:SFRcalc}), removing AGN contamination, for our samples of powerful radio AGNs, but also for our comparison samples of radio-quiet Type-I and Type-II QSOs, ULIRGs, and non-AGN classical elliptical galaxies.

We found that the cool ISM content of SLRGs is enhanced compared to that of non-AGN classical elliptical galaxies, yet below that of ULIRGs (see \S\,\ref{subsec:results_dust}). Galaxy mergers and interactions that are relatively minor in terms of their cool ISM contents are most likely responsible for this enhancement (see \S\,\ref{subsec:discussDistrib}). In contrast, the cool ISM contents of WLRGs associated with FRI-like radio jets are reduced when compared to SLRGs. This is also in contrast with the cool ISM properties of WLRGs/FRIIs, which were found more consistent with that of SLRGs. Therefore, while WLRG/FRIs are mostly triggered by different mechanisms (i.e. direct accretion of the hot gas), at least some WLRG/FRIIs may have been triggered in a similar way to the SLRGs but recently switched off (see \S\,\ref{subsec:discussDistrib}).

No relationships were found between the cool ISM content of our samples of AGNs and the power of the AGN, as traced by \Loiii\ (see \S\,\ref{subsec:AGNvsMgas}). This implies that, while a minimum amount of gas of \Mgas~$\sim~10^8$~\Msun\ is required to trigger the mostly radiatively efficient radio AGNs, other factors must also be important in dictating the AGN power and efficiencies.

Finally, many of the powerful radio AGNs in our samples fall below the MS of \cite{Sargent2014}, implying that their SFRs are lower than expected for galaxies at such stellar masses and redshifts (see \S,\ref{subsec:rejuv}). We argued that this cannot be due to the quenching effect of AGN feedback on star formation, since it would be inconsistent the surprisingly high star formation efficiencies found in SLRGs, which are typical of those observed in star-bursting systems (see \S\,\ref{subsec:AGNvsSFefficiencies}). In fact, we further showed that the location of powerful radio AGNs in the \Rms-\Mstar\ parameter space coincides with that of a population of galaxies believed to be undergoing rejuvenation (see \S,\ref{subsec:rejuv}).

Overall, our results provide strong evidence that the majority of powerful radio AGNs in the local universe are associated with late-time re-triggering of both star formation and AGN activity (rejuvenation), mainly fuelled via galaxy mergers and interactions for SLRGs, and some other mechanisms (e.g. direct accretion of the hot gas) for WLRGs.

\section*{Acknowledgements}
We thank the anonymous referee for the valuable comments which helped improving the quality of the paper. EB, JRM, CT acknowledge STFC grant ST/R000964/1. CRA acknowledges financial support from the EU H2020 research and innovation programme under Marie Sk\l odowska-Curie grant agreement No 860744 (BiD4BESt), from the AEI-MCINN and MCIU under grants RYC-2014-15779, “Feeding and feedback in active galaxies" 
(PID2019-106027GB-C42), “Quantifying the impact of quasar feedback on galaxy evolution” (QSOFEED; EUR2020-112266), and from the Consejería de Econom\' ia, Conocimiento y Empleo del Gobierno de Canarias and the European Regional Development Fund (ERDF) under grant ProID2020010105. This research has made use of the NASA/IPAC Infrared Science Archive (IRSA) and the NASA/IPAC Extragalactic Database (NED) which are operated by the Jet Propulsion Laboratory, California Institute of Technology, under contract with the National Aeronautics and Space Administration. The following packages were used for the data reduction and analysis: MATPLOTLIB \citep{Hunter2007}, ASTROPY \citep{Astropy2018}, NUMPY, SCIPY \citep{scipy}, PANDAS \citep{pandas}, MATH \citep{math}, and NUMBA \citep{numba}.

\section*{Data availability statement}
The archival data of this article were accessed from the IRSA (\url{https://irsa.ipac.caltech.edu/frontpage/}) and the NED (\url{https://ned.ipac.caltech.edu}) databases. Other data (e.g. fluxes from targetted observations) are available in Tables referenced in the text. The new data generated by our analysis have been made available in the supplementary material of this publication.

\bibliographystyle{mnras}
\bibliography{./biblio}

\bsp	
\label{lastpage}
\end{document}